\documentclass[final,5p,times,twocolumn]{elsarticle}

\usepackage{amssymb}
\usepackage{amsmath}
\usepackage{lineno}
\usepackage{subcaption}  
\usepackage{graphicx}    
\usepackage{float}       
\usepackage{booktabs}  
\usepackage{multirow}  
\usepackage{tabularx,siunitx} 
\usepackage{xcolor}     
\usepackage{makecell}
\begin{document}

\begin{frontmatter}

\title{A Multimodal Data Fusion Attention-Empowered Generative Adversarial Network for Real Time 3D Underwater Sound Speed Field Construction} 

\author[label1]{Wei Huang} 
\ead{hw@ouc.edu.cn}
\ead[url]{www.weiwilliamhuang.cn}
\author[label1]{Yuqiang Huang}
\ead{yqhuang@stu.ouc.edu.cn}
\author[label2]{Jixuan Zhou}
\ead{zhoujixuan@whu.edu.cn}
\author[label1]{Hao Zhang\corref{cor1}}
\ead{zhanghao@ouc.edu.cn}
\author[label3]{Tianhe Xu\corref{cor1}}
\ead{thxu@sdu.edu.cn}
\author[label4]{Qian Sun}
\ead{qbcouple@163.com}
\author[label5]{Fang Ji}
\ead{heujifang@163.com}
\cortext[cor1]{Corresponding Author.}
\affiliation[label1]{organization={Faculty of Information Science and Engineering, Ocean University of China},
            addressline={No.238 Songling Road}, 
            city={Qingdao},
            postcode={266100}, 
            state={Shandong},
            country={China}}
\affiliation[label2]{organization={Hanjiang National Laboratory},
	addressline={No.2 Zhangzhidong Road}, 
	city={Wuhan},
	postcode={430079}, 
	state={Hubei},
	country={China}}
\affiliation[label3]{organization={School of Space Science and Technology, Shandong University (Weihai)},
	addressline={No.180 Wenhua West Road}, 
	city={Weihai},
	postcode={264209}, 
	state={Shandong},
	country={China}}
\affiliation[label4]{organization={China Waterborne Transport Research Institute},
	addressline={No.8 Western Tucheng Road, Haidian District}, 
	city={Beijing},
	postcode={100088}, 
	country={China}}
\affiliation[label5]{organization={China Ship Research and Development Academy},
	addressline={No.1 Fengxian East Road, Haidian District}, 
	city={Beijing},
	postcode={100094}, 
	country={China}}
\begin{abstract}
Sound speed profiles (SSPs) are crucial underwater parameters that determine the propagation patterns of acoustic signals, directly influencing the energy efficiency of underwater communication and the accuracy of positioning systems. Conventional techniques for obtaining SSPs, such as matched field processing (MFP), compressive sensing (CS), and deep learning (DL), typically depend on on-site sonar measurements, which impose stringent requirements on the deployment of underwater observation systems. To overcome this limitation and enable high-precision sound speed field reconstruction without the need for on-site underwater data collection, we propose a novel multimodal data-fusion generative adversarial network enhanced with residual attention blocks (MDF-RAGAN). This architecture integrates attention mechanisms to capture global spatial feature correlations effectively, while residual modules are employed to extract subtle perturbations in deep-ocean sound velocity distribution caused by sea surface temperature (SST) variations. Experimental results on a public real-world dataset demonstrate that the proposed model outperforms other state-of-the-art methods, achieving an estimation error of less than 0.3 m/s. Specifically, MDF-RAGAN reduces the root mean square error (RMSE) by nearly half compared to convolutional neural network (CNN) and spatial interpolation (SITP) methods, and attains a 65.8\% RMSE reduction relative to the mean profile method. These results highlight the effectiveness of multi-source fusion and cross-modal attention in enhancing the accuracy and robustness of sound speed profile reconstruction.
\end{abstract}


\begin{highlights}
\item We propose an MDF-RAGAN model to obtain 3D sound speed distribution without on-site data collection.
\item We integrate a residual block to learn the influence of SST on sound velocity.
\item We embed an attention mechanism to learn spatial correlations of sound velocity.
\end{highlights}


\end{frontmatter}

\section{Introduction}
\label{sec1}
The estimation of underwater sound speed profiles (SSPs) has emerged as a significant research focus, owing to its critical role in governing acoustic signal propagation. This, in turn, directly determines the energy efficiency of underwater acoustic communication systems and the precision of underwater positioning  \citep{MUNK1979Tomography,Munk1983Tomography,erol2011survey,Piccolo2019Geoacoustic,Luo2021LocReview}.

Sound speed profiles (SSPs) are popular adopted to describe the distribution of underwater sound speed \cite{Jensen2011COA}, and they are traditionally measured by conductivity, temperature, and depth profiler (CTD) \cite{wang2014CTD,Luo2023CTD,Kirimoto2024CTD} or sound velocity profiler (SVP) \citep{Zhang2022SVP}. The main drawback of instrument measurement methods is their low spatiotemporal resolution. To enhance the efficiency of measuring sound velocity profiles, researchers have developed various inversion methods that utilize sonar observation data by exploiting the dependence of acoustic field properties on the sound speed distribution. These methods can be mainly divided into 3 categories: matching field processing (MFP) \citep{tolstoy1991acoustic,li2010inversion,zhang2012inversion,zhang2015inversion}, compressive sensing (CS) \citep{Bianco2017Dictionary,choo2018compressive}, and deep learning (DL) \citep{stephan1995inverting,Huang2019Collaborating,Huang2023Meta,Huang2024SurveySSP}.

MFP is a method of determining the sound velocity distribution by simulating the SSP and its corresponding theoretical sound field distribution, and matching it with the measured sound field based on ray theory or normal wave theory. However, the MFP method has high algorithm complexity, which reduces its execution efficiency and makes it difficult to continuously construct the sound velocity field.

Considering that ray theory or normal wave theory can only establish a mapping relationship from sound velocity distribution to sound field distribution, scholars have begun to explore how to directly establish a mapping relationship from sound field distribution to sound velocity distribution. The CS technology achieves this process through a feature mapping dictionary, which further improves the efficiency of sound speed estimation compared to MFP. However, the dictionary establishment process introduces linearization approximation \citep{Bianco2017Dictionary}, which sacrifices the accuracy of sound speed construction.

During the last decades, DL has obtained significant achievement in the field of data processing, benefiting from its ability to handle complex nonlinear mapping relationships. The DL-based sound velocity inversion method not only reduces accuracy loss, but also maintains the real-time advantage at the working stage (training can be completed in advance) \citep{Huang2019Collaborating,Huang2023Meta}. However, the aforementioned sound speed profile (SSP) inversion methods are highly dependent on sonar observation data. This imposes stringent requirements on the deployment of sonar observation systems, thereby hindering their capability for large-scale oceanographic coverage.

Recently, to avoid the need for in-situ underwater measurements, researchers have introduced two alternative methods for efficient sound speed estimation. The first approach employs historical sound velocity data to predict current sound speed profiles using time-series forecasting techniques. The second method integrates remotely sensed sea surface temperature (SST) data with historical sound velocity records to reconstruct the sound speed field through spatial interpolation and data fusion techniques. In 2023, Piao et al. proposed an SSP prediction method under internal waves based on long short-term memory (LSTM) neural networks \citep{PIAO2023IntWave}. Lu et al. further developed a hierarchical LSTM (H-LSTM) model that predicts sound speed distribution at different depth layers \citep{Lu2024LSTM}. However, the sound velocity prediction method is only applicable to areas with sufficient historical sound velocity distribution, which can not be satisfied in some ocean areas due to difficulty in measuring SSPs.

To solve this problem, Li et al. \citep{Li2021SOM} and Wu et al. \citep{WU2024SSP} proposed SSP construction methods fusing SST data and historical SSPs, where a self-organized mapping (SOM) neural network was proposed in \citep{Li2021SOM} and a convolutional neural network (CNN) was proposed in \citep{WU2024SSP}. However, the approaches proposed by \citep{Li2021SOM} and \citep{WU2024SSP} are primarily confined to modeling adjacent spatial dependencies. Their failure to capture long-range global relationships thus imposes significant limitations on model accuracy.

To achieve high-precision estimation of the sound speed profile (SSP) in a given sea area without relying on in situ measurements, we propose a multi-modal data-fusion generative adversarial network incorporating a residual attention block (MDF-RAGAN). This model is designed to enhance the capture of global spatial feature correlations through attention mechanisms, while residual modules are employed to extract fine-grained perturbations in deep-ocean sound velocity distribution caused by SST variations. The contributions of our work are concluded as follows:
\begin{itemize}
\item To efficiently obtain the sound velocity distribution in a given ocean area, we propose an MDF-RAGAN model that integrates SST data with historical SSPs, thereby eliminating the reliance on on-site underwater measurements. In this framework, the generator learns a mapping from the fused multimodal inputs to the target sound speed distribution, while the discriminator constrains the spatiotemporal consistency and physical plausibility of the generated profiles. 
\item To enhance the accuracy of underwater SSP reconstruction, we introduce a residual attention mechanism into a multi-modal data-fusion GAN. The residual blocks enable the model to capture subtle variations in sound velocity caused by SST changes, while the attention module improves its ability to model long-range spatial dependencies and regional correlations within large-scale oceanic environments.
\item To evaluate the feasibility of the proposed MDF-RAGAN model, the performance was tested on real sampled SSP and SST data. Results show that MDF-RAGAN surpasses the state-of-the-art (SOTA) models in the field of sound velocity distribution estimation.	
\end{itemize}
The rest part of this paper is organized as follows. Section \ref{sec2} provides a detailed description of the proposed MDF-RAGAN framework. In Section \ref{sec3}, the model is thoroughly evaluated and its performance is discussed based on experiments conducted on a real-world open dataset. Finally, Section \ref{sec4} concludes the paper and outlines potential directions for future research.

\section{Sound Speed Estimation by MDF-RAGAN}
\label{sec2}
\subsection{Overview of MDF-RAGAN Framework}
In this study, we seek to reconstruct the SSP at a target location using remote sensing SST data and the historical SSP distributions. This task presents two major challenges: first, the spatial coordinates, SST values, and SSP measurements constitute heterogeneous modal data, requiring effective alignment and fusion within a unified feature representation; second, the SSPs exhibit highly nonlinear dependencies on environmental factors such as depth, temperature, and salinity, demanding a model with strong expressive power to capture these complex relationships.

\begin{figure}[!htbp]
	\centering
	\includegraphics[width=0.5\textwidth]{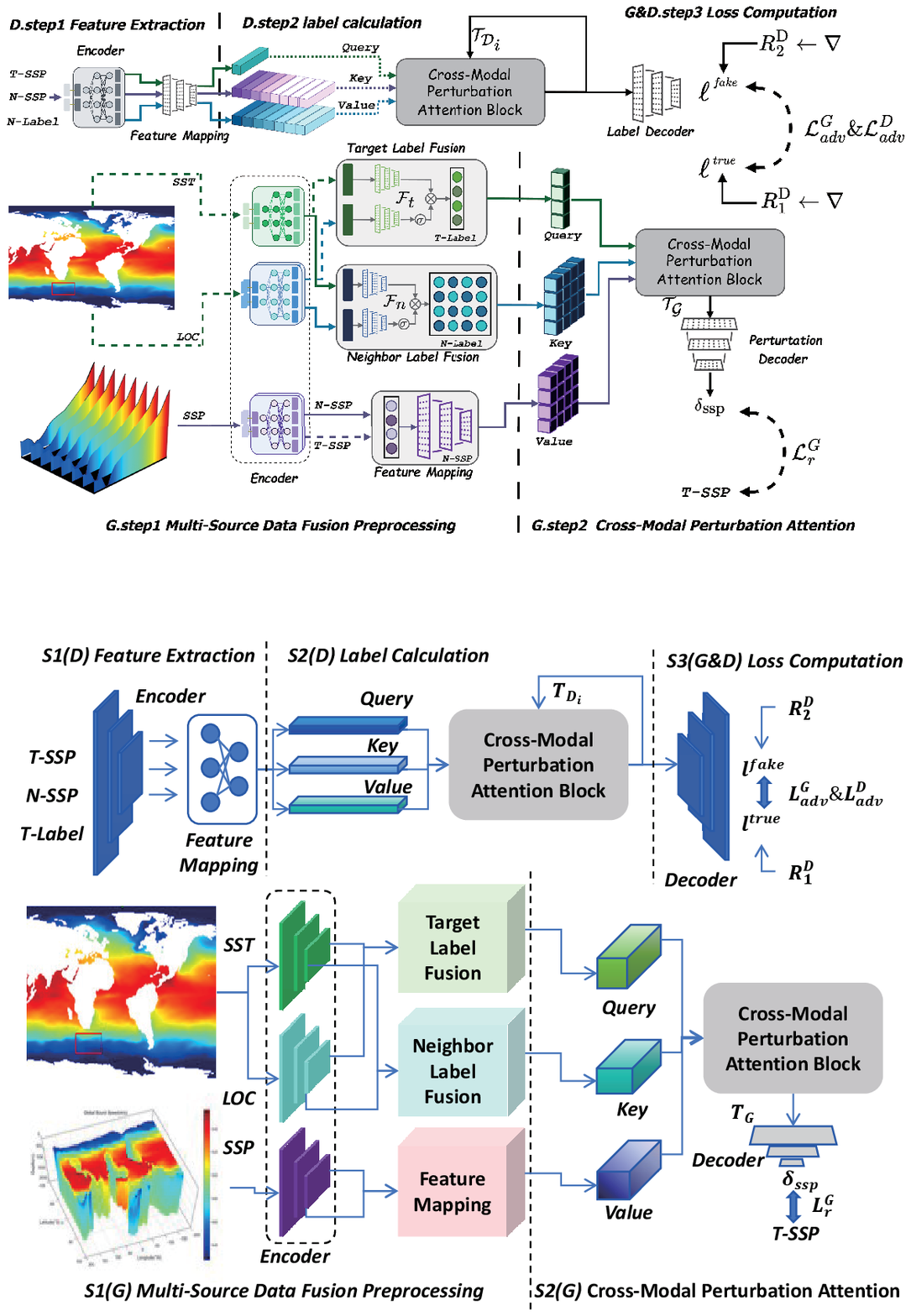}
	\caption{Framework of SSP estimation based on the MDF-RAGAN}
	\label{fig01}
\end{figure}

To address these challenges, we propose MDF-RAGAN, a novel framework comprising three key components, as illustrated in Fig.~\ref{fig01}. First, we map three types of heterogeneous input modalities, namely, the geographic coordinates of the target point and its surrounding grid, SST values, and the average profile of historical SSPs, into a unified feature space through multi-source data fusion preprocessing. Subsequently, a cross-modal perturbation attention mechanism is employed to capture spatial correlations within sound velocity distributions. Finally, an end-to-end optimization strategy is established by integrating generative adversarial training with multi-task regression loss, ensuring robust prediction performance even with limited samples.

In physical oceanography, underwater sound-speed profiles are governed by temperature, salinity, and pressure, with temperature exerting the dominant control in the upper ocean. As the most accessible and spatially continuous thermal observable, sea surface temperature (SST) provides a key boundary condition on mixed-layer temperature and upper-ocean stratification, and previous studies have shown that basin-scale sound-speed structures can be reconstructed from sea-surface variables such as SST and sea surface height (SSH), underscoring the tight coupling between surface thermodynamics and subsurface acoustics \citep{liu2024research}.

Under climatological and regional regimes, the vertical temperature profile \(T(z)\) therefore exhibits a statistically stable dependence on SST, geographic location, and season, which has been extensively exploited in Argo-based empirical reconstructions and oceanographic climatologies. Building on this physical prior, MDF-RAGAN integrates SST, historical SSPs, and geographic coordinates so that SST constrains the surface and mixed-layer state, historical SSPs encode the large-scale background structure, and their joint representation guides the reconstruction of full-depth SSPs in data-sparse regions.

The input of the model are coordinates, SST and historical SSPs. Let the set of SSPs at the reference points surrounding a given latitude-longitude coordinate be denoted as
$\mathbf{R}_{ssp}^{n}\in \mathbb{R}^{N\times D}$, where $N$ denotes the number of surrounding reference points, and $D$ represents the depth dimension. The geographic coordinates (longitude $\lambda$, latitude $\phi$) and the SST obtained by remote sensing at each reference point are denoted by $\mathbf{R}_{loc}^{n}\in \mathbb{R}^{N\times 2}, \quad \mathbf{R}_{sst}^{n}\in \mathbb{R}^{N\times 1}$. The geographic coordinates and the SST at the target point are denoted as $\mathbf{R}_{loc}^{t}\in\mathbb R^{1\times2}, \mathbf{R}_{sst}^{t}\in\mathbb R^{1\times1}$.

The final output from the generator network is a small perturbation $\mathbf{\delta}_{ssp}$ at the target location. Therefore, the estimated SSP $\mathbf{\hat{s}}_{ssp}$ at the target point can be expressed as the sum of the average SSP of surrounding reference points and the generated perturbation:
\begin{equation}
\mathbf{\hat{s}}_{ssp}=\frac{1}{N}\sum_{n=1}^N{\mathbf{R}_{ssp}^{n}+\mathbf{\delta} _{ssp}}
\end{equation}

This decomposition strategy guides the generator network to explicitly focus on capturing variations at the target location. By leveraging the general trend information derived from surrounding reference points, the estimated SSP is expressed as the sum of the average profile of nearby references and a residual perturbation term. The generator is designed to learn this perturbation term, which compensates for the limitations of the mean profile, particularly under complex marine environmental conditions, thereby significantly improving prediction accuracy. Furthermore, this decomposition simplifies the training process, allowing the network to concentrate on modeling deep nonlinear deviations rather than reconstructing the background mean field.

\begin{figure}[!htbp]
	\centering
	\includegraphics[width=0.5\textwidth]{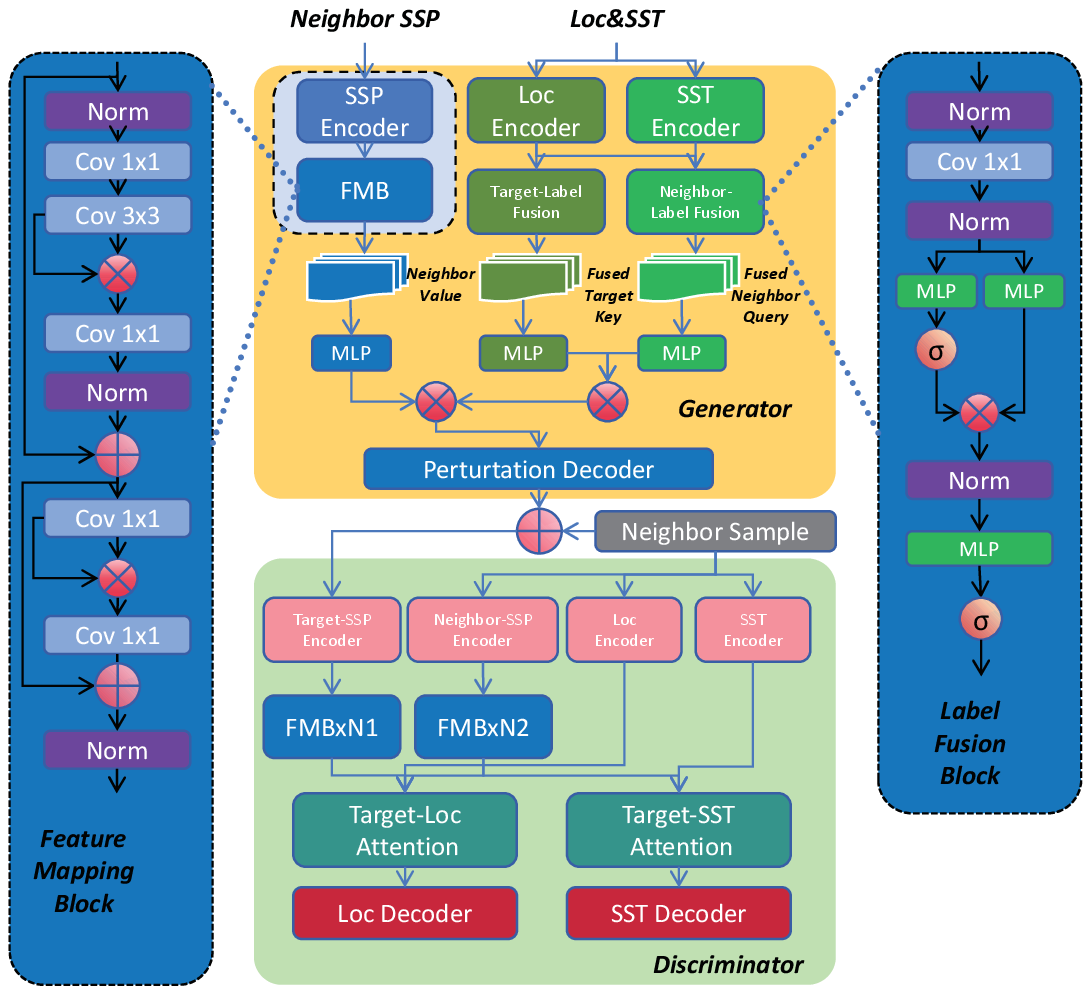}
	\caption{The proposed MDF-RAGAN model for SSP estimation.}
	\label{fig02}
\end{figure}
The detailed model of MDF-RAGAN is given in Fig.~\ref{fig02}. In the following section, we detail the architectures of the generator and discriminator, along with the optimization strategy.

\subsection{Generator}
\label{sec2-2}
The primary objective of the generator is to map fused multi-source features into detailed SSP perturbations at the target location, thereby reconstructing ocean temperature-induced variations in the sound velocity distribution. The entire network operates sequentially through three stages: feature encoding and fusion, cross-modal cross-attention, and perturbation decoding.

\subsubsection{Multi-Source Data Fusion}
\label{sec2-2-1}
The SSP exhibits complex nonlinear behavior due to variations in ocean depth, temperature, and salinity. Geographic coordinates help identify the thermohaline structure of a given ocean region, while SST distribution directly influences the surface thermohaline gradient, significantly affecting the upper ocean sound speed distribution. To accurately estimate the SSP at target points, it is essential to effectively align and fuse multi-source data, such as geographic coordinates and SST, within a unified feature space, thereby leveraging their complementary physical relationships.

\paragraph{Label Fusion Block}
The label fusion block (LFB) is designed to adaptively integrate label information from both the target point and its surrounding reference points. Specifically, the latitude-longitude coordinates and remote sensed SST data are first embedded into a feature space of dimension $d_r$ via respective encoding pathways. Each feature stream is processed through a linear transformation followed by layer normalization:
\begin{equation}
\mathbf{z}_{loc}^{t}=W_{loc}\mathbf{R}_{loc}^{t}+b_{loc},
\end{equation}
\begin{equation}
\mathbf{z}_{sst}^{t}=W_{sst}\mathbf{R}_{sst}^{t}+b_{sst},
\end{equation}
where $W_{loc}\in\mathbb R^{2\times d_r}$ and $W_{sst}\in\mathbb R^{1\times d_r}$ are the linear transformation matrices that map these two types of inputs to the $d_r$-dimensional feature space, with the corresponding bias terms being $b_{loc}\in\mathbb R^{d_r}$ and $b_{sst}\in\mathbb R^{d_r}$. Thus, we can obtain the $d_r$ dimensional embedding feature vectors for surface remote sensing temperature and coordinate points.

The same feature transformations are applied to the corresponding reference point data, mapping the reference point features into a consistent dimension. The aligned labels are concatenated along the sample dimension to form the fusion matrix $\mathbf{Z}^{r}\in \mathbb{R}^{N\times d_r}$, which is then processed through a learnable fusion transformation with instance normalization and learnable scaling/shifting parameters. After the transformation, we divide the fusion matrix $\mathbf{\bar{Z}}^r$ into two parts $\mathbf{\bar{Z}}_{A}^{r}, \mathbf{\bar{Z}}_{B}^{r} \in \mathbb{R}^{N\times d_r}$ and feed them to the linear gating unit GLU:
\begin{equation}
\mathbf{Z}_L=\mathbf{\bar{Z}}_{A}^{r}\odot \sigma \left( \mathbf{\bar{Z}}_{B}^{r} \right).
\end{equation}

This enables adaptive weighting across multi-modal labels by dynamically adjusting the relative importance of coordinates and temperature in SSP predictions based on the geographic and thermal conditions of different ocean regions.
\paragraph{Feature Mapping Block}
Since detailed SSPs vary with depth due to scattering layers and thermoclines, the feature mapping block (FMB) employs local context modeling to amplify these critical physical features. Given the SSP features at the reference points $ \mathbf{Z}_{ssp}^{r}\in \mathbb{R}^{1\times N\times d_r} $, the FMB initially extracts fine-grained, depth-wise features through depthwise separable convolutions:
\begin{equation}
\mathbf{\dot{Z}}_{ssp}^{r}=\text{DSConv}\left( \mathbf{Z}_{ssp}^{r} \right).
\end{equation}

The depthwise separable convolution first applies depthwise convolution to each channel independently, followed by a pointwise convolution to fuse channel information. In deep-sea environments, abrupt variations in salinity across adjacent depth levels may amplify gradients during training. To mitigate potential gradient explosion introduced by convolutional operations alone, we further apply a SimpleGate activation module after convolution:
\begin{equation}
SG\left( \mathbf{\dot{Z}}_{ssp}^{r}\right) =\mathbf{\dot{Z}}_{ssp}^{r}\odot \mathbf{\dot{Z}}_{ssp}^{r}.
\end{equation}

These variations are captured, and in turn, enhanced features are constructed using residual linking and scaling factors:
\begin{equation}
\mathbf{Z}_{fmb}^{r,n}=\mathbf{Z}_{ssp}^{r}+\xi \cdot SG\left( \mathbf{\dot{Z}}_{ssp}^{r} \right),
\end{equation}
where the weight $\xi$ is a learnable scaling factor used to balance the relative contributions of the original depth features $\mathbf{Z}_{ssp}^r$ and the enhanced features $SG\left( \mathbf{\dot{Z}}_{ssp}^{r} \right)$ in the residual link. Through training, the model automatically adjusts the size of $\xi$: when $\xi$ is large, it emphasizes the local nonlinear changes captured by the depthwise separable convolution and SimpleGate; when $\xi$ is small or even close to 0, it focuses more on preserving the overall trend of the SSP.

This design not only improves the sensitivity of the model to transition zones such as the thermocline and scattering layer, but also effectively emphasises local deviations while preserving the overall trend, which helps the generator to accurately simulate the detailed variations in the SSP at the target point.

By synergistically integrating the LFB and FMB, the module aligns multimodal labels with deep spatial context features and enhances them concurrently. This process yields rich, structured, and physically meaningful initial features for MDF-RAGAN, establishing a foundation for robust and precise adversarial learning.

\subsubsection{Cross-Modal Perturbation Attention Block}
After feature encoding and fusion, we obtain the SSP features from the reference point and the fused label embeddings of the target and reference points after processing by the label fusion module $ \mathbf{Z}_{L}^{t}\in \mathbb{R}^{1\times d_r},\mathbf{Z}_{L}^{r,n}\in \mathbb{R}^{N\times d_r} $. The goal is to accurately predict the fine-grained perturbation of the SSP at the target point based on the label information. To this end, we propose the Cross-Modal Perturbation Attention Block module to capture the cross-modal dependencies and guide the perturbation modeling process.

Specifically, we treat the fused label embedding of the target point as the query $ \mathbf{Q} $, the fused label embeddings of the reference points as the key $ \mathbf{K} $, and the reference SSP features of the surrounding points as the value $ \mathbf{V} $. These tensors are then passed through learned linear projections to obtain the initial query, key, and value matrices:
\begin{equation}
Q^{\left( 0 \right)}=\mathbf{Z}_{L}^{t}W_{Q}^{\left( 0 \right)}+b_{Q}^{\left( 0 \right)},K^{\left( 0 \right)}=\mathbf{Z}_{L}^{r,n}W_{K}^{\left( 0 \right)}+b_{K}^{\left( 0 \right)},V^{\left( 0 \right)}=\mathbf{Z}_{fmb}^{r,n}W_{V}^{\left( 0 \right)}+b_{V}^{\left( 0 \right)}.
\end{equation}

To ensure that longitudinal depth-wise positional information is not lost during the computation of attention, an aligned positional encoding is required. That is, a learnable positional encoding $ PE\in \mathbb{R}^{D\times d} $ is superimposed to each vector, allowing the model to correctly distinguish the natural order of sound speed profiles at different depths. The cross-modal attention mechanism requires a number of computations, of which the $ l $-th layer of computation can be expressed as:

\begin{equation}
X^{\left( l \right)}=\frac{Q^{\left( l-1 \right)}\left( K^{\left( l-1 \right)} \right) ^{\mathbf{T}}}{\sqrt{d}},m_{i}^{\left( l \right)}=\max X_{ij}^{\left( l \right)},
\end{equation}

\begin{equation}
A_{ij}^{\left( l \right)}=\frac{\exp \left( X_{ij}^{\left( l \right)}-m_{i}^{\left( l \right)} \right)}{\sum_{j^{\prime}}{\exp \left( X_{ij^{\prime}}^{\left( l \right)}-m_{i}^{\left( l \right)} \right)}},O^{\left( l \right)}=A^{\left( l \right)}V^{\left( l-1 \right)}.
\end{equation}

In the above calculation, $d$ denotes the dimension size of the query and key vectors, and $l$ denotes the index of the attention layer. The elements of the unnormalised score matrix $X^{(l)}\in\mathbb R^{N\times N}$ represent the similarity between the $i$-th query and the $j$-th key. $m_i^{(l)}=\max_j X_{ij}^{(l)}$ takes the maximum value of each row to stabilise the values before the Softmax operation. The normalised attention weights $A_{ij}^{(l)}$ represent the model's attention intensity for each key, and the final output matrix $O^{(l)}$ is the weighted sum whose dimensions are consistent with $V^{(l-1)}$. The subscripts $i$ and $j$ correspond to the query position and key/value position in the sequence (or depth layer), respectively.

Considering that multiple calculations may produce exploding gradients, here we subtract the maximum value of each row when calculating Softmax, which can effectively avoid the exploding-gradient problem during training caused by too large or too small values. Temperature, salinity, and pressure often vary in magnitude from one reference point to another, and with this adaptive logit normalization , the model is able to propagate multi-point information into each depth level with more balanced weights, thus simulating a weighted contribution effect similar to that of interpolation.

The attention output $ O^{l} $ is processed with FFN activation function and layer normalization to increase non-linear representational capacity:

\begin{equation}
H^{\left( l \right)}=W_{2}^{\left( l \right)}\,\text{GELU}\bigl( W_{1}^{\left( l \right)}\,\text{LN}\left( O^{\left( l \right)} \right) +b_{1}^{\left( l \right)} \bigr) +b_{2}^{\left( l \right)}.
\end{equation}

The residual connection adds $ O^{l} $ to the layer input, stabilizing deep-layer information propagation and maintaining smooth gradient flow. After the $ l $-th cross-modal attention layer, the resulting feature $ H^{l} $ is fed into a perturbation decoding head to predict the fine-grained SSP perturbation through a two-layer fully connected network:

\begin{equation}
\delta _{ssp}=W_{dec2}\,\text{GELU}\bigl( W_{dec1}H^{\left( L \right)}+b_{dec1} \bigr) +b_{dec2}.
\end{equation}

This perturbation is added to the mean SSP of the reference points to generate the final prediction $ \mathbf{\hat{s}}_{ssp} $. In summary, the cross-modal attention mechanism in the generator combined with the numerical stabilization strategy mimics the balanced distribution of the contributions from each detection point in the acoustic-profile interpolation, and exploits the nonlinear capability of the deep network to capture the complex sound velocity structures such as thermocline and scattering layer, thus achieving a fine-grained prediction of the SSP perturbation at the target point.

\subsection{Discriminator}
\label{sec2-3}

The core of the discriminator is to distinguish between real and generated SSP samples, and to provide more informative gradient signals to the generator through multi-task regression. The discriminator accepts two types of target SSP inputs: the real samples $ \mathbf{s}_{ssp}^{real}\in \mathbb{R}^D $  and the generated samples $ \mathbf{s}_{ssp}^{fake}=\frac{1}{N}\sum_{i=1}^N{\mathbf{R}_{ssp,i}^{n}+\delta _{ssp}}\in \mathbb{R}^D $ output by the generator, which, together with the real reference point samples, constitute the inputs to the discriminator network:
\begin{equation}
X_{in}=\left[ \mathbf{s}_{ssp};\mathbf{R}_{ssp}^{n};\mathbf{R}_{sst}^{n};\mathbf{R}_{loc}^{n} \right] 
\end{equation}

Similarly, we encode the input target SSP $ \mathbf{s}_{ssp} $ of unknown authenticity and the real reference point SSP $ \mathbf{R}_{ssp}^{n} $ respectively, and also encode the coordinate information $ \mathbf{R}_{loc}^{n}$  and SST obtained by remote sensing $ \mathbf{R}_{sst}^{n} $ of the reference point, aligned to the same feature space. After the correct SSP is extracted by FMB, the sequence of SSPs and the labeling information jointly participate in feature learning collaboratively.

Here we use the coordinates of the target point and the SST as the output of the discriminator, which is used as an alternative to the traditional way of training the discriminator in terms of realism. The input tensor is stacked with the same multi-layer stack of cross-modal attention blocks and position-wise feed-forward networks used in the generator, where the Query is the SSP feature vector of the target point $\mathbf{Z}_{ssp}^{t}$, the Key is the SSP feature vector of the reference point $\mathbf{Z}_{ssp}^{r,n}$, and the Value consists of the feature vectors encoded by the reference point coordinates $\mathbf{Z}_{loc}^{r,n}$ and the SST $\mathbf{Z}_{sst}^{r,n}$, respectively. The discriminator applies this attention mechanism with identical numerical stabilization as described in Section~\ref{sec2-2-1} to enforce physical consistency between the generated SSP and the associated environmental conditions.


This numerical-stabilization trick subtracts the row-wise maximum logit before the soft-max, mitigating exploding or vanishing gradients and reducing the risk of mode collapse during adversarial training.

After obtaining the output features $ H_{loc}^{\left( l \right)}\in \mathbb{R}^{1\times d_r} $ and $ H_{ssp}^{\left( l \right)}\in \mathbb{R}^{1\times d_r} $ , we split into two branches, where the coordinate regression branch maps the latitude and longitude of the target point, and the SST regression branch predicts the sensing SST through a similar fully connected layer:
\begin{equation}
\mathbf{\hat{t}}_{loc}=W_{loc}H_{loc}^{\left( l \right)}+b_{loc},
\end{equation}
\begin{equation}
\mathbf{\hat{t}}_{sst}=W_{sst}H_{sst}^{\left( l \right)}+b_{sst}.
\end{equation}

In contrast to conventional GAN discriminators that output a single real/fake logit, the discriminator in MDF-RAGAN is formulated as a multi-task regressor that explicitly predicts the SST and geographic coordinates associated with each candidate SSP. This design yields richer and smoother gradient signals than a one-bit decision, since the regression logits encode physically meaningful environmental constraints tied to the underlying oceanographic state. By forcing the discriminator to recover these variables, the adversarial learning process enforces cross-modal physical consistency and suppresses solutions that are acoustically plausible but environmentally unrealistic. As a result, the discriminator serves not only as a real/fake judge but also as a validator of oceanographic coherence, which stabilises training and enhances the fidelity of the reconstructed sound-speed perturbations in complex marine environments.

\subsection{Optimization Strategy}
\label{sec2-4}

In this section, the loss function as well as different training strategies are designed mainly for the generator and the discriminator, so as to balance the stability of the game-based training as well as the accuracy of the SSP perturbation reconstruction. The generator loss mainly consists of the  relativistic Softplus GAN loss and the centroid SSP reconstruction loss. Specifically, the discriminator produces two sequences of realism scores for real target-point SSP samples:
\begin{equation}
\ell _{i}^{real}=D\left( \mathbf{s}_{ssp}^{real};\mathbf{R}_{ssp}^{n};\mathbf{R}_{sst}^{n};\mathbf{R}_{loc}^{n} \right),
\end{equation}
\begin{equation}
\ell _{i}^{fake}=D\left( \mathbf{s}_{ssp}^{fake};\mathbf{R}_{ssp}^{n};\mathbf{R}_{sst}^{n};\mathbf{R}_{loc}^{n} \right).
\end{equation}
Then the generator against loss is defined as:
\begin{equation}
\mathcal{L}_{adv}^{G}=-\frac{1}{B}\sum_{i=1}^B{\text{Softplus}}\bigl( \ell _{i}^{fake}-\ell _{i}^{real} \bigr).
\end{equation}
The minus sign is to allow the generator to approximate the true SSP distribution by maximizing the fake-sample score, thus ensuring that the SSP data at the target point can be correctly reconstructed from the label data. The generator also introduces a root mean square error (RMSE) loss of the SSP at the center-point in order to approximate the true value:
\begin{equation}
\mathcal{L}_{r}^{G}=\frac{1}{B}\sum_{i=1}^B{\sqrt{\frac{1}{D}\sum_{j=1}^D{\bigl( s_{ij}^{real}-s_{ij}^{fake} \bigr) ^2}}},
\end{equation}
where $ s_{ij}^{real} $ and $ s_{ij}^{fake} $ denote the true and predicted values of the $ i $-th sample at depth $ j $, respectively, $B$ is the batch size, and $D$ is the number of depth points. Combining the two losses, the total loss of the generator can be expressed as:
\begin{equation}
\mathcal{L}_G=\mathcal{L}_{adv}^{G}+\eta \mathcal{L}_{r}^{G}
\end{equation}

The loss function of the discriminator consists of an adversarial true-false branch as well as a zero-centered gradient penalty. The adversarial branching uses the relative discriminative power of the generated versus true samples along the Softplus form of the metric:
\begin{equation}
\mathcal{L}_{adv}^{D}=\frac{1}{B}\sum_{i=1}^B{\text{Softplus}}\bigl( \ell _{i}^{fake}-\ell _{i}^{real} \bigr).
\end{equation}

A zero-centered gradient penalty is also introduced in order to limit training instability due to too large a gradient, and the R1 and R2 terms are computed separately for real and generated samples:

\begin{equation}
R_{1}^{D}=\mathbb{E}_{\mathbf{s}^{real}}\bigl[ |\nabla _{\mathbf{s}}D\left( \mathbf{s}^{real},\mathbf{R}_{ssp}^{t} \right) |_{2}^{2} \bigr],\label{eq23}
\end{equation}
\begin{equation}
R_{2}^{D}=\mathbb{E}_{\mathbf{s}^{fake}}\bigl[ |\nabla _{\mathbf{s}}D\left( \mathbf{s}^{fake},\mathbf{R}_{ssp}^{t} \right) |_{2}^{2} \bigr].\label{eq24}
\end{equation}
In Eqs. \eqref{eq23} and \eqref{eq24}, the zero-centred gradient penalty terms $R_1^D$ and $R_2^D$ measure the gradient magnitude of the discriminator on the real sample distribution $\mathbf{s}^{real}$ and the generated sample distribution $\mathbf{s}^{fake}$, respectively. Specifically, $\mathbb{E}_{\mathbf{s}^{real}}$ and $\mathbb{E}_{\mathbf{s}^{fake}}$ denote the expectation over the distributions corresponding to the real and fake samples; and $\nabla_{\mathbf{s}} D(\mathbf{s},\mathbf{R}_{ssp}^{t})$ is the gradient of the discriminator's output $D$ with respect to its input sound speed profile $\mathbf{s}$, where the conditional input $\mathbf{R}_{ssp}^{t}$ represents the reference sound speed profiles.

Eventually we can get the total loss function representation of the discriminator:
\begin{equation}
\mathcal{L}_D=\mathcal{L}_{adv}^{D}+\,\eta _1R_{1}^{D}+\eta _2\,R_{2}^{D}.
\end{equation}

In order to guarantee the stability and convergence effect of MDF-RAGAN in the early stage of adversarial training, we adopt a staged training strategy. In the first stage, we only optimise the generator parameters $\theta_G$, while the discriminator parameters $\theta_D$ are kept frozen and not updated. The generator learns an initial mapping from the multi-source fused label embeddings to the SSP perturbations at the target point. A linear learning-rate warm-up is applied, and the network parameters are updated solely with the generator loss $\mathcal{L}_G$. In the second stage, training switches to adversarial learning: network parameters are alternately updated with the generator loss $\mathcal{L}_G$ and the discriminator loss $\mathcal{L}_D$, enabling joint optimization of both networks.

The learning rate $\eta _t$ can be scheduled using the cosine annealing learning rate:
\begin{equation}
\eta _t=\eta _{\min}+\frac{1}{2}\left( \eta _{\max}-\eta _{\min} \right) \Bigl( 1+\cos \bigl( \left. \pi \tfrac{t}{T} \bigr) \Bigr) \right. 
\end{equation}
where $T$ is the total number of epochs and $ t $ is the current epoch, $\eta_{\max}$ is the maximum initial learning rate used at the end of warm-up and $\eta_{\min}$ is the minimum learning rate at the end of annealing. Such a staged training strategy effectively ensures that the generator obtains a robust initial mapping without the disturbance first, and obtains a more accurate prediction of the sound speed perturbation in the subsequent adversarial phase by alternating game-like training.

\section{Results and Discussions}
\label{sec3}
\subsection{Experimental Setting}
\label{sec3-1}
\subsubsection{Dataset}
\label{sec3-1-1}

In this study, an experimental dataset was constructed on the sea area (gridded $1^{\circ} \times 1^{\circ}$ latitude and longitude) from $59.5^{\circ}$ to $39.5^{\circ}$ S and $0.5^{\circ}$ to $38.5^{\circ}$ E, combining Argo observations and remotely sensed information.

SSP data were taken from the GDCSM\_Argo gridded Argo dataset \citep{GDCSM}, and the raw observation profiles covered 0-1976 m depth with 58 scattered points, which were linearly interpolated depth-by-depth to 1 m intervals to generate 1977-dimensional high-resolution SSP vectors. SST data were obtained from the NOAA OISST monthly average product \citep{SST2021}, spatially mapped and aligned to an Argo grid to obtain the current month's SST values for each grid point.

In order to make full use of the spatial neighborhood information, each $1^{\circ} \times 1^{\circ}$ grid is taken as the center point, and a $3 \times 3$ sliding window is formed with its surrounding eight adjacent grid points. The interpolated SSP at the center is used as the estimation target, and the coordinates, SST and SSP of the remaining eight grid points are processed by the LFB with the FMB to output the multimodal input features used for generator perturbation estimation.

In terms of timing, to assess the impact of different historical data lengths on model performance, the training set is divided into three time windows: January 2004-December 2020, January 2011-December 2020, and January 2016-December 2020; the test set uniformly uses data from January 2021 to June 2023. During spatial sampling, the training samples are extracted at 3° intervals along the latitude and longitude directions to ensure that there is no data leakage during testing; the test samples are offset by 3° on the basis of the training grid to achieve spatial non-overlapping coverage, ensuring that the training and testing areas are spatially complementary and without overlap.

Leveraging the above multi-source fusion and sliding-window sampling strategy, we construct a high-quality dataset containing thousands of spatiotemporally matched samples. This dataset provides a solid foundation for validating the model's generalization ability under different historical windows, unseen spatial regions, and multimodal feature settings.

\subsubsection{Performance Evaluation Metrics}
\label{sec3-1-2}
To evaluate the accuracy and robustness of the model for estimating SSPs at target points, the RMSE was adopted in this paper. Let the test set contain $ N_{te} $ samples, the true SSP at the target point for the first $ n $ sample is noted as $ \mathbf{s}_{n}^{t} $ and the generator predicts the profile as $ \mathbf{\hat{s}}_{n}^{t} $, then the RMSE denotes the average of $L_2$ distance:
\begin{equation}
	\text{RMSE}=\sqrt{\frac{1}{N_{te}}\sum_{i=1}^{N_{te}}{\lVert \mathbf{\hat{s}}_{n}^{t}-\mathbf{s}_{n}^{t} \rVert _2}}
\end{equation}
This indicator visualizes the magnitude of the absolute error between the generated profile and the true observation.

\subsubsection{Parameter Initialization}
\label{sec3-1-3}

All experiments were conducted on a Linux server equipped with a single NVIDIA RTX 3090 (24 GB) and an Intel Xeon Gold 6330 CPU. Training and inference were implemented in PyTorch.

The generator and discriminator share an embedding dimension (hidden size) of 384, with a dropout rate and attention-dropout rate of 0.05. Each network uses [2, 2] Transformer blocks connected via residual paths.

Training hyper-parameters were set as follows: batch size 128, epochs 196, learning rate of the generator $lr_g = 4 \times 10^{-4}$, learning rate of the discriminator $lr_d = 5 \times 10^{-4}$. The learning rate was linearly warmed up for the first 20 epochs and then decayed with cosine annealing, then the minimum learning rate was $lr_{min} = 1 \times 10^{-7}$, and the weight decay was set to be $1 \times 10^{-3}$. Data loading employs 8 data-loader workers.

\subsection{Baselines}
\label{sec4-2}

In order to comprehensively validate the performance enhancement of the proposed MDF-RAGAN model in SSP estimation, three types of baseline methods are selected in this study: averaged profile based on historical statistics, spatial interpolation (SITP), and convolutional neural network (CNN) \citep{WU2024SSP}. Quantitative results show that MDF-RAGAN, by effectively fusing physical priors and modelling nonlinear interactions, consistently outperforms these baselines.

\subsubsection{Mean Profile}
\label{sec4-2-1}

In the simplest historical statistical baseline, we compute the mean value of the sound velocity for each depth layer in the training set and use that average profile as an estimation for the corresponding depth for all test points. Let there be a total of $ N $ sample points around the target point, the mean profile estimation equation is:
\begin{equation}
\mathbf{\hat{s}}_{}^{avr}=\frac{1}{N}\sum_{i=1}^N{\mathbf{s}_{i}^{ngb}}
\end{equation}

\subsubsection{Spatial Interpolation}
\label{sec4-2-2}

We mainly use the inverse distance weighting method to assign different weights to the estimated values for each depth layer by considering the spatial distance between the target point and its surrounding consistent sample points. Specifically, for the location $ \left( x,y \right) $ , the interpolated SSP is:
\begin{equation}
\mathbf{\hat{s}}_{}^{IDW}\left( x,y \right) =\frac{\sum_{n=1}^N{\varpi _n\mathbf{s}_n}}{\sum_{n=1}^N{\varpi _n}},\varpi _n=\frac{1}{D_{n}^{p}}
\end{equation}
where $ D_n=\sqrt{\left( x-x_n \right) ^2+\left( y-y_n \right) ^2} $ , is the in-plane distance between the target point and the location of the first $n$  known profile, usually taken as a power exponent $ p=2 $ . The SITP method weights historical observations within a geographic neighborhood, and can capture the effects of thermohaline gradients at spatially adjacent points to some extent.

\subsubsection{Attention-Assisted CNN Model}
\label{sec4-2-3}

As a deep learning baseline, this study implements Att-CNN model for comparing the performance of deep networks in SSP tasks. The Att-CNN model is an improved version of the CNN architecture proposed in \citep{WU2024SSP}, incorporating attention mechanisms to capture long-range dependencies in the depth dimension \citep{wu2026attention}. Compared with the SOM-based method \citep{Xu2025SOM} which reconstructs SSP from remote sensing data using self-organizing maps, the Att-CNN demonstrates superior performance and represents the current state-of-the-art approach for SSP estimation. The model takes the center and eight profile vectors, corresponding coordinates and SST of its 3×3 neighborhood grid as multi-channel inputs, and extracts local features in the depth dimension through three-layer one-dimensional convolution (convolution kernel width of 3, channel numbers of 64, 128 and 256 in turn), ReLU activation and maximum pooling; subsequently, the perturbation value of the center is outputted by the regression of the two-layer fully-connected network. During the training process, the CNN is also optimized using the RMSE as the loss function.

\subsection{Accuracy Performance}
\label{sec4-3}
\subsubsection{Influence of Training Data Length}
\begin{figure}[!htbp]
	\centering
	\begin{subfigure}[b]{0.8\linewidth}
		\includegraphics[width=\linewidth]{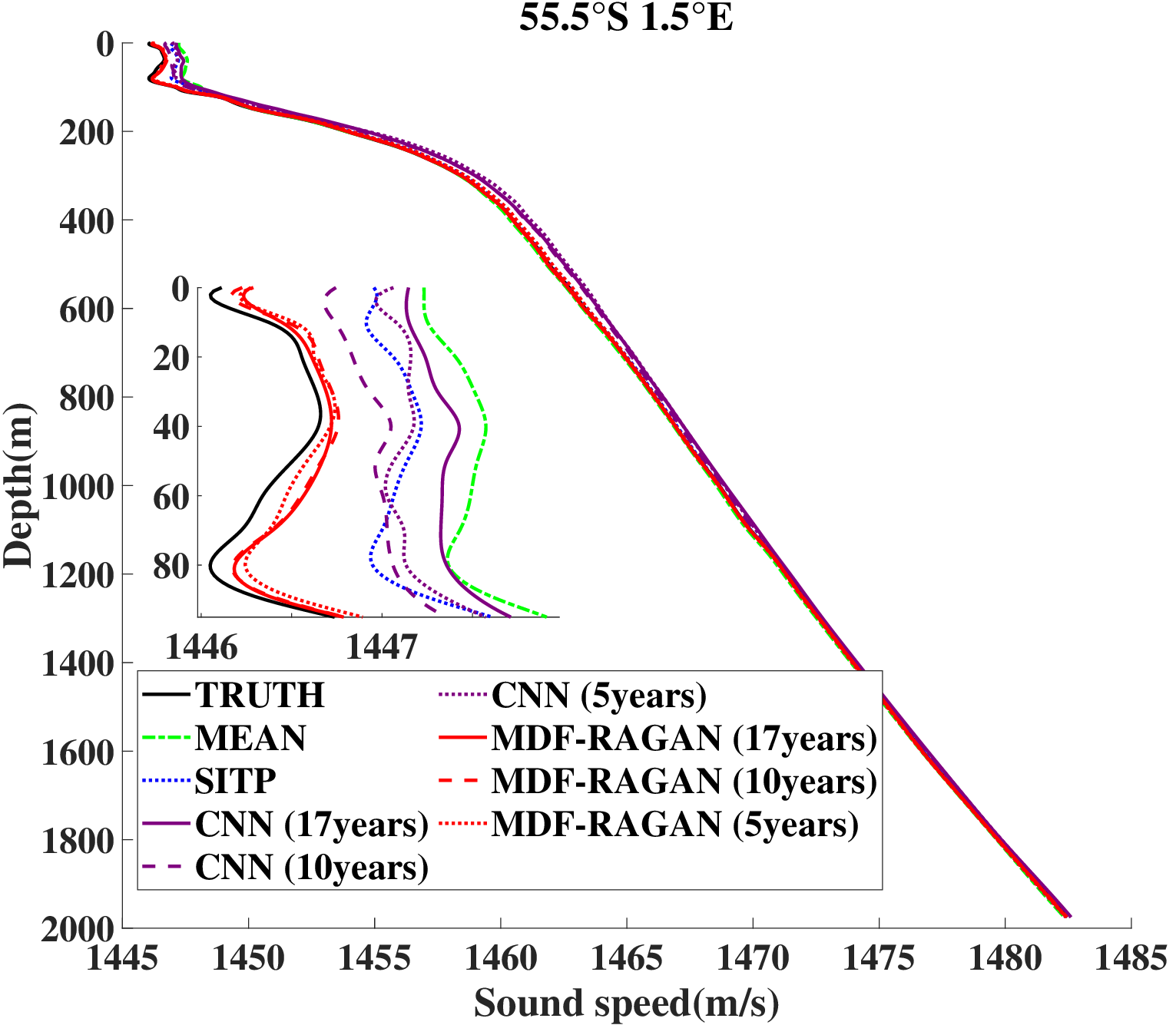}
		\caption{Location 1: ($55.5^{\circ}$S, $1.5^{\circ}$E), 1975m}
	\end{subfigure}
	\begin{subfigure}[b]{0.8\linewidth}
		\includegraphics[width=\linewidth]{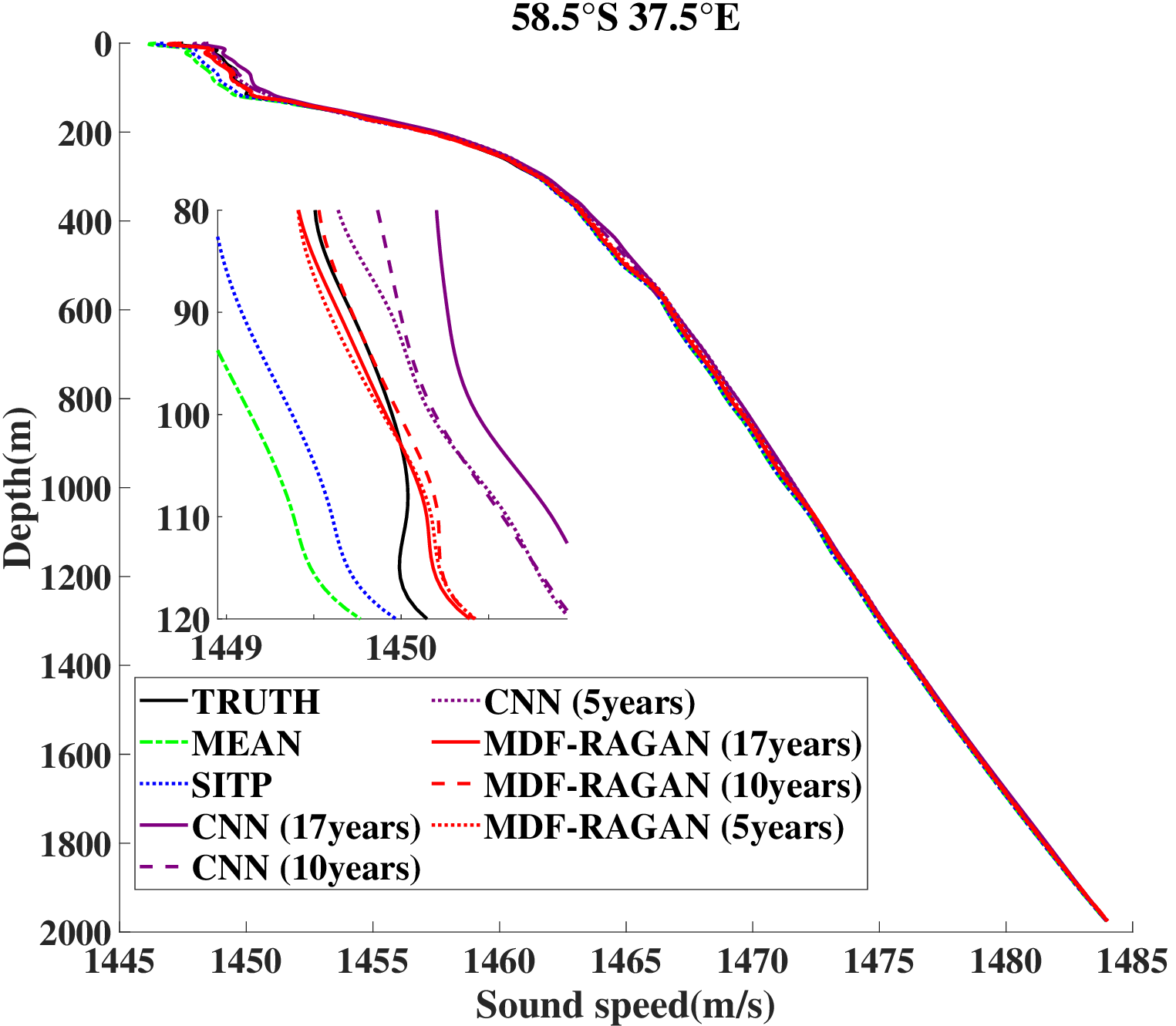}
		\caption{Location 2: ($55.5^{\circ}$S, $37.5^{\circ}$E), 1975m}
	\end{subfigure}
	\caption{
		Comparison of sound speed profile predictions at different locations and depths using models trained with different historical data lengths. Each subfigure compares the predictions of MDF-RAGAN, CNN, SITP, and MEAN using training data of 5, 10, and 17 years.
	}
	\label{fig03}
\end{figure}

To evaluate the influence of training data length on model accuracy performance, Fig.~\ref{fig03} compares the estimation performance of models trained using three historical windows of 5 years (2016-2020), 10 years (2011-2020) and 17 years (2004-2020) on the same sea profile. As shown in the figure, the window length from 3 to 10 years extends the RMSE of each method in a decreasing trend, but the MDF-RAGAN can achieve an RMSE of about 0.18 m/s under the condition of the shortest 3 years of data, and with the increase of the training years its error further converges to about 0.14 m/s, without any significant decrease, which indicates that the model already has a strong representation ability for a small number of historical samples.

\begin{table*}[htbp]
	\centering
	\fontsize{8pt}{10pt}\selectfont
	\caption{Comparison of RMSE with different training data lengths}
	\label{tab01}
	\begin{tabular}{ccccccccc}
		\toprule
		\multirow{2}{*}{Method} & \multirow{2}{*}{Location} & \multicolumn{3}{c}{1000 m} & \multicolumn{3}{c}{1975 m} \\
		\cmidrule(lr){3-5} \cmidrule(lr){6-8}
		& & 5 years & 10 years & 17 years & 5 years & 10 years & 17 years \\
		\midrule
		\multirow{4}{*}{MDF-RAGAN} & ($58.5^{\circ}$S, $19.5^{\circ}$E) & 0.187 & 0.138 & 0.125 & 0.140 & 0.105 & 0.095 \\
		& ($58.5^{\circ}$S, $26.5^{\circ}$E) & 0.168 & 0.151 & 0.128 & 0.128 & 0.114 & 0.100 \\
		& ($55.5^{\circ}$S, $22.5^{\circ}$E) & 0.226 & 0.176 & 0.176 & 0.168 & 0.133 & 0.131 \\
		& ($52.5^{\circ}$S, $22.5^{\circ}$E) & 0.266 & 0.242 & 0.245 & 0.199 & 0.181 & 0.182 \\
		\midrule
		\multirow{4}{*}{Att-CNN} & ($58.5^{\circ}$S, $19.5^{\circ}$E) & 0.455 & 0.460 & 0.615 & 0.336 & 0.341 & 0.457 \\
		& ($58.5^{\circ}$S, $26.5^{\circ}$E) & 0.545 & 0.445 & 0.572 & 0.408 & 0.326 & 0.427 \\
		& ($55.5^{\circ}$S, $22.5^{\circ}$E) & 0.541 & 0.482 & 0.661 & 0.412 & 0.366 & 0.501 \\
		& ($52.5^{\circ}$S, $22.5^{\circ}$E) & 0.648 & 0.542 & 0.613 & 0.487 & 0.410 & 0.465 \\
		\midrule
		\multirow{4}{*}{SITP} & ($58.5^{\circ}$S, $19.5^{\circ}$E) & 0.437 & 0.437 & 0.437 & 0.319 & 0.319 & 0.319 \\
		& ($58.5^{\circ}$S, $26.5^{\circ}$E) & 0.194 & 0.194 & 0.194 & 0.151 & 0.151 & 0.151 \\
		& ($55.5^{\circ}$S, $22.5^{\circ}$E) & 0.383 & 0.383 & 0.383 & 0.315 & 0.315 & 0.315 \\
		& ($52.5^{\circ}$S, $22.5^{\circ}$E) & 0.385 & 0.385 & 0.385 & 0.357 & 0.357 & 0.357 \\
		\midrule
		\multirow{4}{*}{MEAN} & ($58.5^{\circ}$S, $19.5^{\circ}$E) & 0.427 & 0.427 & 0.427 & 0.318 & 0.318 & 0.318 \\
		& ($58.5^{\circ}$S, $26.5^{\circ}$E) & 0.219 & 0.219 & 0.219 & 0.164 & 0.164 & 0.164 \\
		& ($55.5^{\circ}$S, $22.5^{\circ}$E) & 0.457 & 0.457 & 0.457 & 0.373 & 0.373 & 0.373 \\
		& ($52.5^{\circ}$S, $22.5^{\circ}$E) & 0.503 & 0.503 & 0.503 & 0.484 & 0.484 & 0.484 \\
		\bottomrule
	\end{tabular}
\end{table*}

In addition, Table~\ref{tab01} compares the RMSE of MDF-RAGAN trained on 5, 10 and 17 years of historical data with the corresponding CNN, as well as SITP with the historical mean (MEAN) method, at four different depths (200 m, 400 m, 1000 m, 1975 m) and four typical shallow and deep sea coordinate points. Overall, the RMSE of MDF-RAGAN shows a continuous decrease at all depths and locations as the length of the training data expands from 5 to 17 years: for example, the RMSE of MDF-RAGAN at a depth of 200 m (point 1: ($58.5^{\circ}$S, $19.5^{\circ}$E)) decreases from 0.332 m/s to 0.236 m/s; at the same location at a depth of 1000 m, its error is more than halved from 0.187 m/s to 0.095 m/s. This shows that the longer historical window allows the model to learn a more adequate pattern of thermohaline change, especially in the deep ocean region, where the error decreases more significantly. Further comparing the different depths, it can be seen that the initial RMSE of MDF-RAGAN is higher than that of the deep sea (1000 m, 1975 m) in the shallow sea region (200 m and 400 m), but with the increase of the training years, its error rapidly approaches the deep-sea level, and stays in the low range of 0.10-0.20 m/s in all depths at the 17-year window, showing a weak sensitivity to the depth, while the CNN and traditional methods, although performing better in the deep sea, consistently maintain a high amplitude error interval of 0.20-0.95 m/s at shallow depths. In the following experiments, the training data length will be 17 years.

\begin{table}[htbp]
	\centering
	\fontsize{8pt}{10pt}\selectfont 
	\caption{RMSE comparison of different methods at 1975 m depth across 10 typical locations (data from 2021.01-2023.06) (m/s)}
	\label{tab02}
	\begin{tabularx}{\linewidth}{@{\extracolsep{\fill}}ccccc@{}}  
		\toprule
		\textbf{Location} & \textbf{MDF-RAGAN} & \textbf{Att-CNN} & \textbf{MEAN} & \textbf{SITP}  \\
		\midrule
		($58.5^{\circ}$S, $1.5^{\circ}$E) & 0.067 & 0.344 & 0.312 & 0.221 \\
		($49.5^{\circ}$S, $13.5^{\circ}$E) & 0.119 & 0.270 & 0.303 & 0.205 \\
		($49.5^{\circ}$S, $16.5^{\circ}$E) & 0.134 & 0.290 & 0.292 & 0.199 \\
		($49.5^{\circ}$S, $19.5^{\circ}$E) & 0.215 & 0.353 & 0.289 & 0.212 \\
		($49.5^{\circ}$S, $22.5^{\circ}$E) & 0.188 & 0.371 & 0.346 & 0.288 \\
		($49.5^{\circ}$S, $31.5^{\circ}$E) & 0.095 & 0.220 & 0.442 & 0.375 \\
		($46.5^{\circ}$S, $10.5^{\circ}$E) & 0.108 & 0.216 & 0.547 & 0.556 \\
		($46.5^{\circ}$S, $37.5^{\circ}$E) & 0.154 & 0.229 & 0.258 & 0.202 \\
		($43.5^{\circ}$S, $22.5^{\circ}$E) & 0.218 & 0.276 & 0.391 & 0.375 \\
		($43.5^{\circ}$S, $37.5^{\circ}$E) & 0.173 & 0.324 & 0.738 & 0.508 \\
		\midrule
		\textbf{Average} & \textbf{0.147} & \textbf{0.289} & \textbf{0.392} & \textbf{0.314} \\
		\bottomrule
	\end{tabularx}
\end{table}

\subsubsection{Influence of Coverage Density}

To evaluate the sensitivity of SSP reconstruction to the spatial coverage density of historical profiles, we enlarge the original \(3\times3\) neighbourhood to a sparser \(5\times5\) window while still selecting nine reference profiles. This configuration increases the average spacing between reference points, weakens local spatial correlations, and thus provides a more challenging sparse-coverage scenario. As summarised in Table~\ref{tab:coverage-depth}, all methods exhibit increased RMSEs under this setting, indicating that denser historical sampling is beneficial for accurate SSP reconstruction.

The degree of performance degradation, however, varies substantially across models. MEAN and SITP show clear error inflation at all depths, and the Att-CNN baseline also suffers noticeable accuracy loss, especially within the upper 400 m where thermocline variability is strongest. By contrast, MDF-RAGAN consistently achieves the lowest RMSE across all depth layers and locations: even with sparse neighbourhoods, its integrated-profile errors over 0--1000 m and 0--1975 m remain at 0.459 and 0.336 m/s, compared with 0.653/0.482 m/s for Att-CNN and 0.964/0.745 m/s for SITP (Table~\ref{tab:coverage-depth}), and it maintains a clear advantage at representative stations along the meridional transect (Table~\ref{tab:coverage-spatial}).

\begin{table}[htbp]
	\centering
	\fontsize{8pt}{10pt}\selectfont
	\caption{RMSE (m/s) of different methods at various depths under the sparse-coverage setting, characterising the effect of vertical coverage density of historical SSPs.}
	\label{tab:coverage-depth}
	\begin{tabular}{lcccc}
		\toprule
		\textbf{Depth (m)} & \textbf{MDF-RAGAN} & \textbf{Att-CNN} & \textbf{MEAN} & \textbf{SITP} \\
		\midrule
		200  & 0.632 & 0.952 & 1.591 & 1.391 \\
		400  & 0.575 & 0.844 & 1.458 & 1.241 \\
		1000 & 0.459 & 0.653 & 1.168 & 0.964 \\
		1975 & 0.336 & 0.482 & 0.907 & 0.745 \\
		\bottomrule
	\end{tabular}
\end{table}

To further examine the impact of spatial coverage density, we analyse a meridional transect where historical sampling gradually decreases from high to lower latitudes, spanning \((57.5^{\circ}\mathrm{S},\,22.5^{\circ}\mathrm{E})\) to \((42.5^{\circ}\mathrm{S},\,2.5^{\circ}\mathrm{E})\). Table~\ref{tab:coverage-spatial} summarises the RMSEs of each method at 200 m, 400 m, and 1000 m for four representative locations. As coverage becomes sparser equatorward, all methods show error growth, but MDF-RAGAN degrades more gently than Att-CNN, MEAN, and SITP, and keeps RMSE values below 0.8 m/s at all depths and stations.

\begin{table*}[htbp]
	\centering
	\fontsize{8pt}{10pt}\selectfont
	\caption{RMSE (m/s) of different methods at representative locations with varying historical coverage density.}
	\label{tab:coverage-spatial}
	{
	\setlength{\tabcolsep}{2pt}
	\begin{tabular}{lcccccccccccc}
		\toprule
		& \multicolumn{3}{c}{($57.5^{\circ}$S, $22.5^{\circ}$E)} & \multicolumn{3}{c}{($52.5^{\circ}$S, $22.5^{\circ}$E)} & \multicolumn{3}{c}{($47.5^{\circ}$S, $22.5^{\circ}$E)} & \multicolumn{3}{c}{($42.5^{\circ}$S, $2.5^{\circ}$E)} \\
		\cmidrule(lr){2-4} \cmidrule(lr){5-7} \cmidrule(lr){8-10} \cmidrule(lr){11-13}
		\textbf{Method} & 200 m & 400 m & 1000 m & 200 m & 400 m & 1000 m & 200 m & 400 m & 1000 m & 200 m & 400 m & 1000 m \\
		\midrule
		MDF-RAGAN & 0.477 & 0.390 & 0.274 & 0.702 & 0.631 & 0.521 & 0.720 & 0.674 & 0.521 & 0.556 & 0.535 & 0.412 \\
		Att-CNN       & 1.001 & 0.801 & 0.561 & 0.916 & 0.803 & 0.622 & 1.052 & 0.934 & 0.740 & 0.863 & 0.751 & 0.562 \\
		MEAN      & 3.229 & 2.553 & 1.945 & 2.087 & 1.548 & 1.082 & 1.114 & 1.259 & 1.015 & 0.768 & 1.125 & 0.881 \\
		SITP      & 2.819 & 2.256 & 1.788 & 1.493 & 1.106 & 0.790 & 0.857 & 0.892 & 0.681 & 0.900 & 0.890 & 0.659 \\
		\bottomrule
	\end{tabular}
	}
\end{table*}

The robustness of MDF-RAGAN under sparse coverage mainly stems from two aspects. First, the residual perturbation formulation allows the generator to preserve the large-scale background structure while only correcting fine-scale deviations, even when reference profiles are widely spaced. Second, the cross-modal perturbation attention aggregates long-range, SST and location-consistent information from historical profiles, suppressing the impact of locally missing observations. As a result, MDF-RAGAN exhibits the smallest performance degradation among all compared methods and remains the most accurate approach under sparse-sampling conditions, which is crucial for large-scale ocean acoustic applications in data-poor regions.

\subsubsection{Accuracy in Deep and Shallow Water}
To evaluate the accuracy performance in estimating the SSP, we compare the results of MDF-RAGAN with mean profile, SITP and CNN on 10 typical latitude and longitude coordinate points (30 months of data from 2021.01-2023.06) as shown in Table~\ref{tab02}. The results show that the RMSE of the historical average MEAN method fluctuates around 0.25$\sim$0.73 m/s, and the RMSE of SITP fluctuates around 0.19$\sim$0.55 m/s. While CNN compresses the results to around 0.21$\sim$0.37 m/s, the MDF-RAGAN proposed in this paper focuses on 0.06$\sim$0.21 m/s, and the mean value is only 0.147 m/s, which is 62\% lower than the traditional mean method, and more than 53\% compared with the spatial interpolation method. Even compared with CNN, the performance is improved by 49\%. The table fully demonstrates the accuracy and stability of MDF-RAGAN for estimation in a large-scale sound velocity gradient environment.

\begin{figure*}[htbp]
	\centering
	\begin{subfigure}[b]{0.32\textwidth}
		\includegraphics[width=\textwidth]{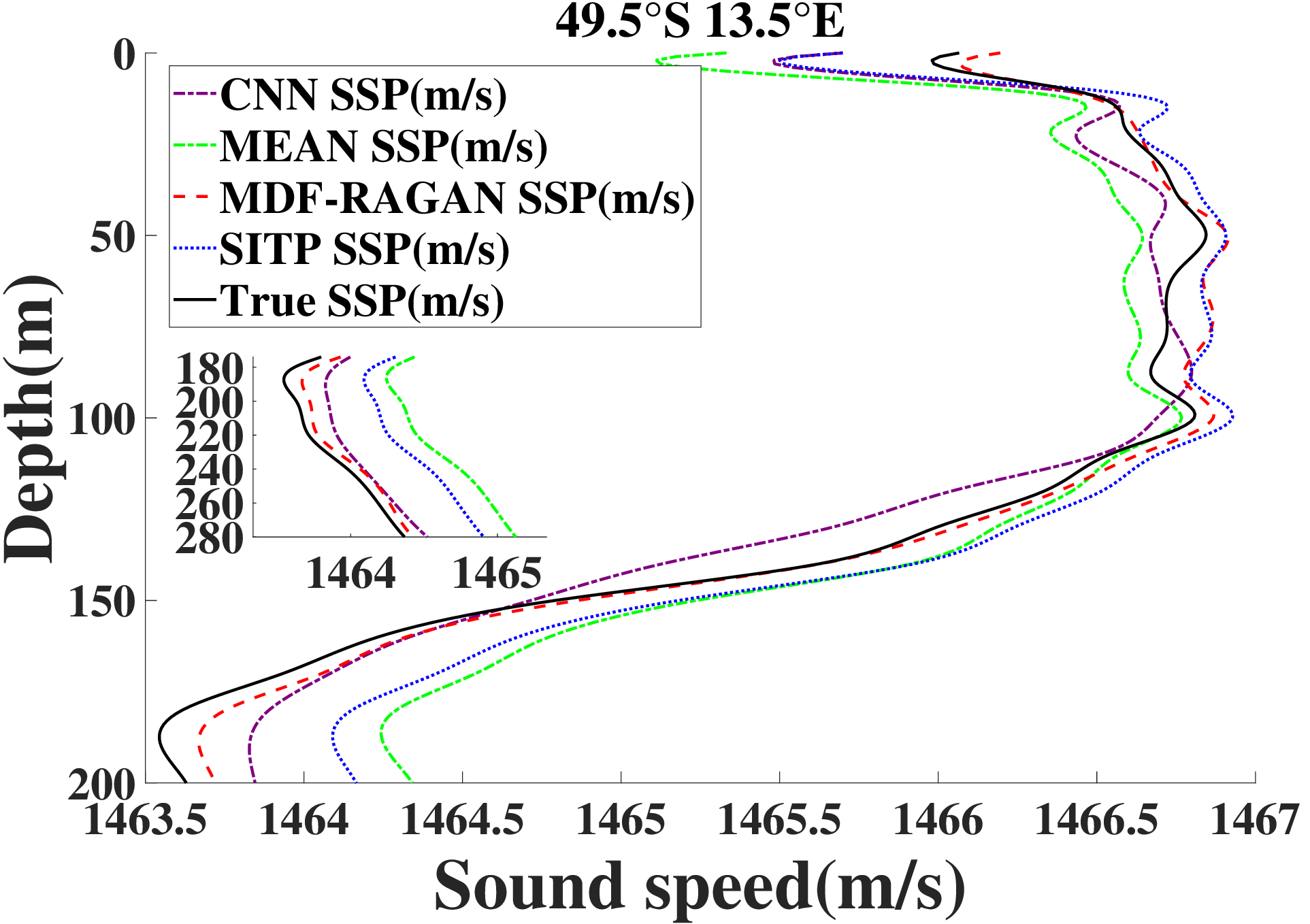}
		\caption{($49.5^{\circ}$S, $13.5^{\circ}$E), 200m}
		\label{fig:ssp-comp-a}
	\end{subfigure}
	\begin{subfigure}[b]{0.32\textwidth}
		\includegraphics[width=\textwidth]{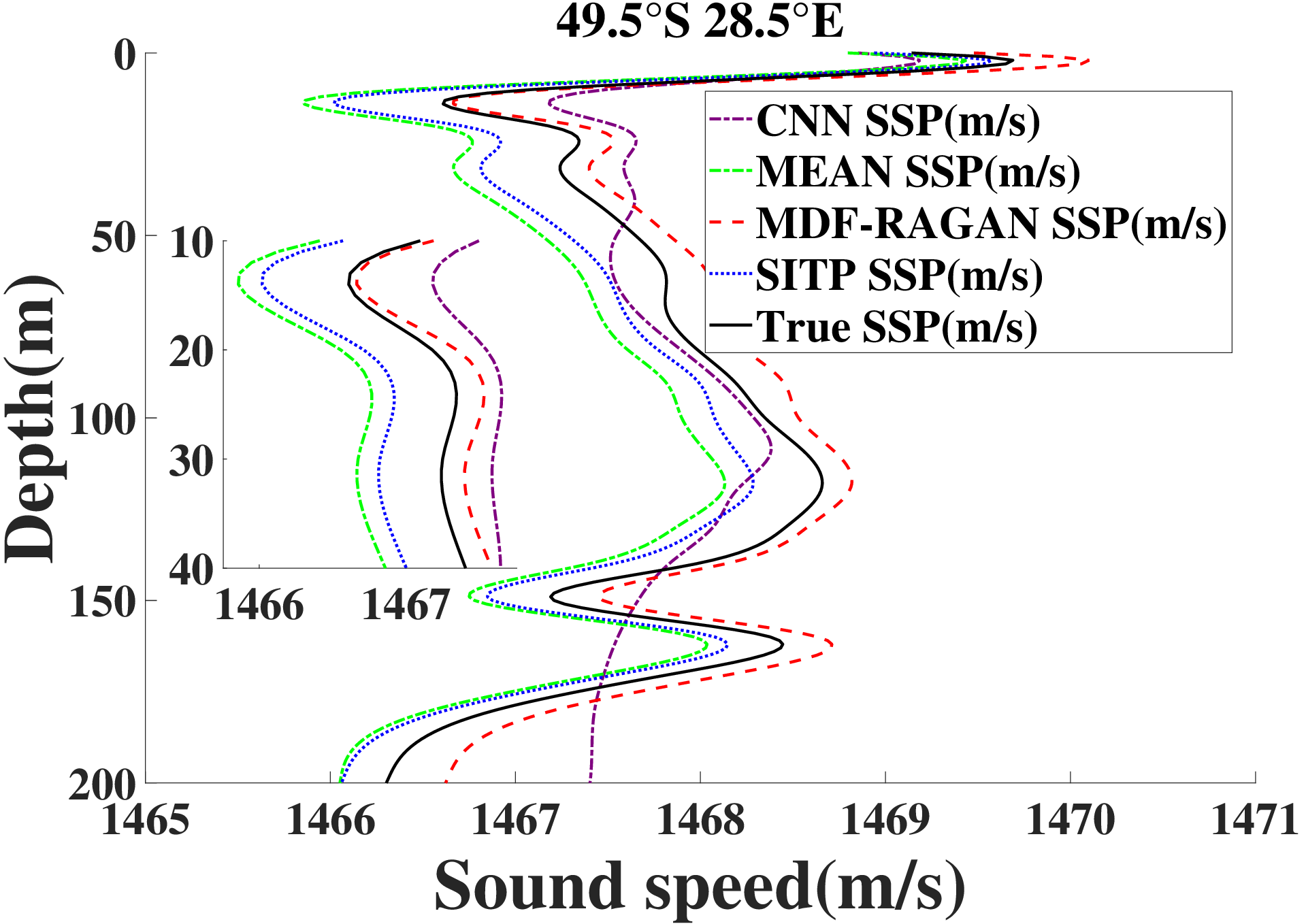}
		\caption{($49.5^{\circ}$S, $28.5^{\circ}$E), 200m}
		\label{fig:ssp-comp-b}
	\end{subfigure}
	\begin{subfigure}[b]{0.32\textwidth}
		\includegraphics[width=\textwidth]{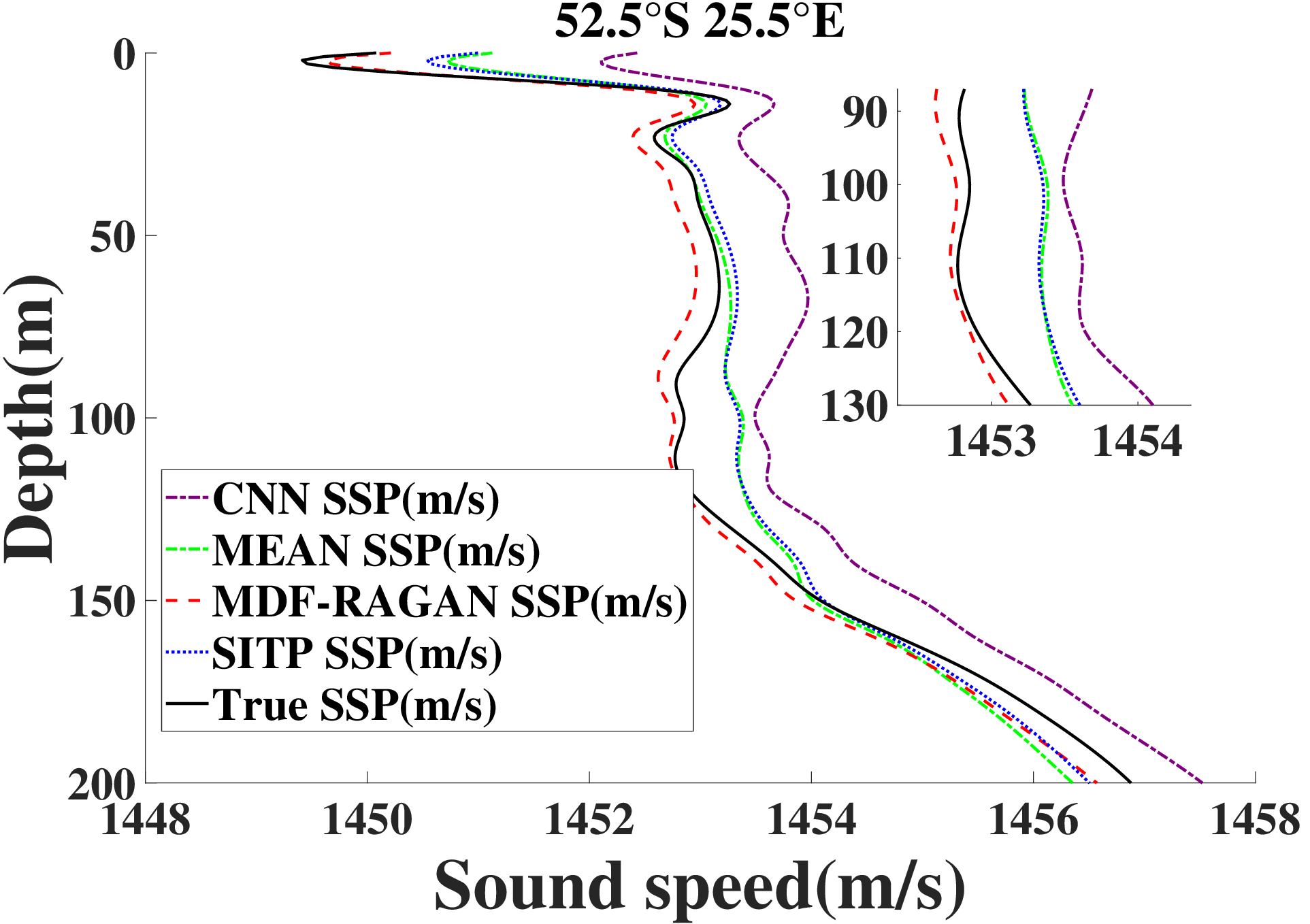}
		\caption{($52.5^{\circ}$S, $25.5^{\circ}$E), 200m}
		\label{fig:ssp-comp-c}
	\end{subfigure}
	
	\begin{subfigure}[b]{0.32\textwidth}
		\includegraphics[width=\textwidth]{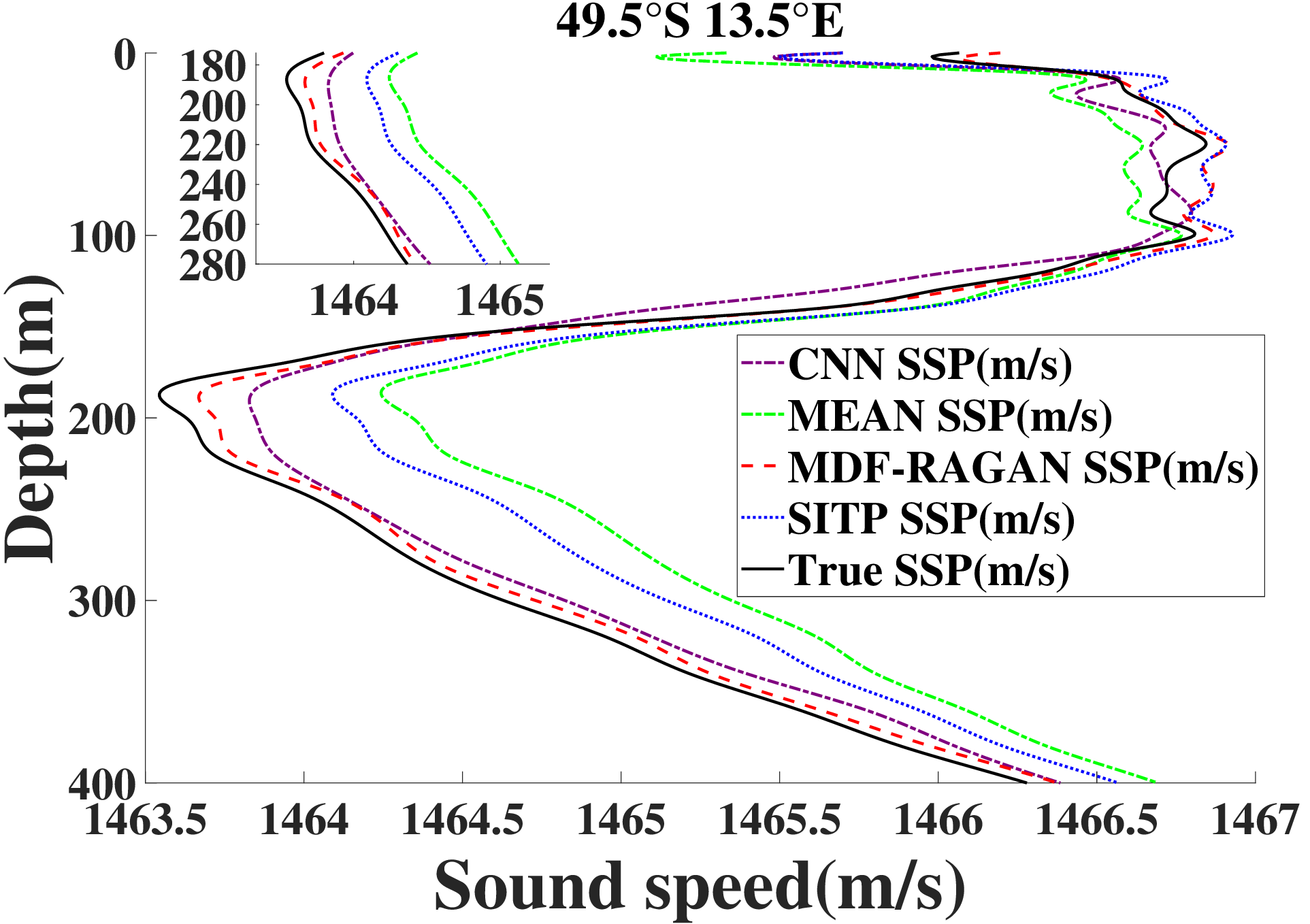}
		\caption{($49.5^{\circ}$S, $13.5^{\circ}$E), 400m}
		\label{fig:ssp-comp-d}
	\end{subfigure}
	\begin{subfigure}[b]{0.32\textwidth}
		\includegraphics[width=\textwidth]{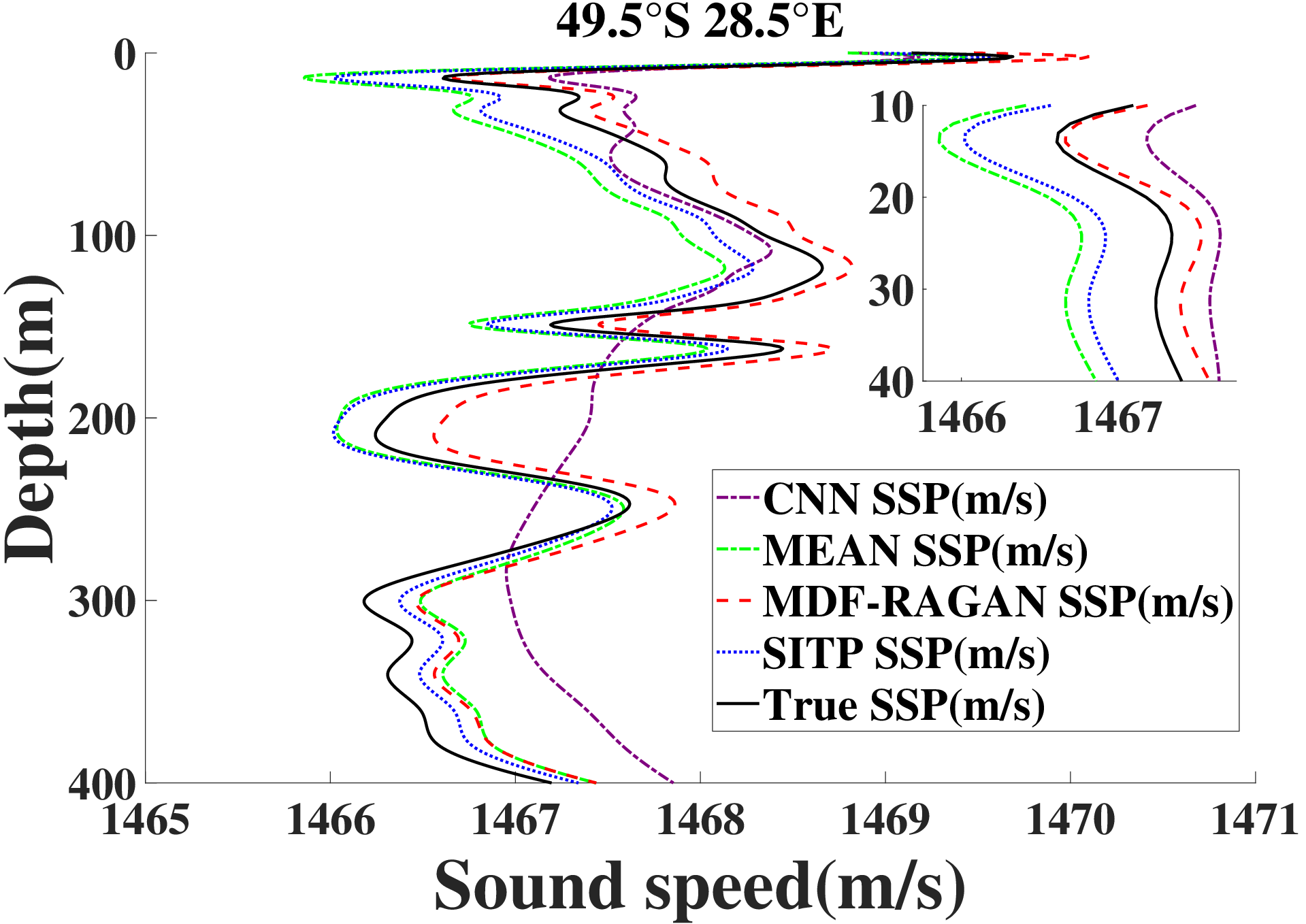}
		\caption{($49.5^{\circ}$S, $28.5^{\circ}$E), 400m}
		\label{fig:ssp-comp-e}
	\end{subfigure}
	\begin{subfigure}[b]{0.32\textwidth}
		\includegraphics[width=\textwidth]{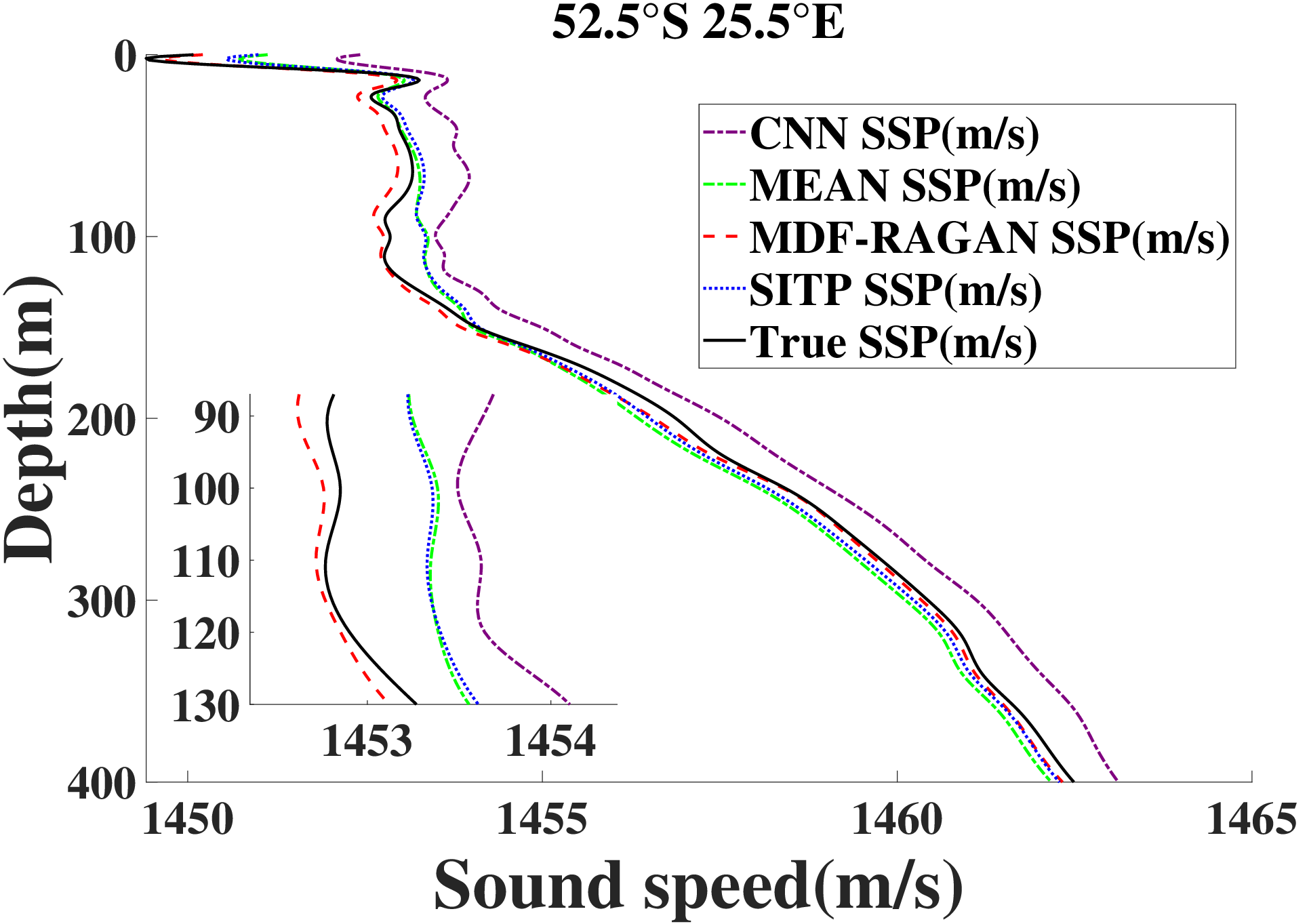}
		\caption{($52.5^{\circ}$S, $25.5^{\circ}$E), 400m}
		\label{fig:ssp-comp-f}
	\end{subfigure}
	
	\begin{subfigure}[b]{0.32\textwidth}
		\includegraphics[width=\textwidth]{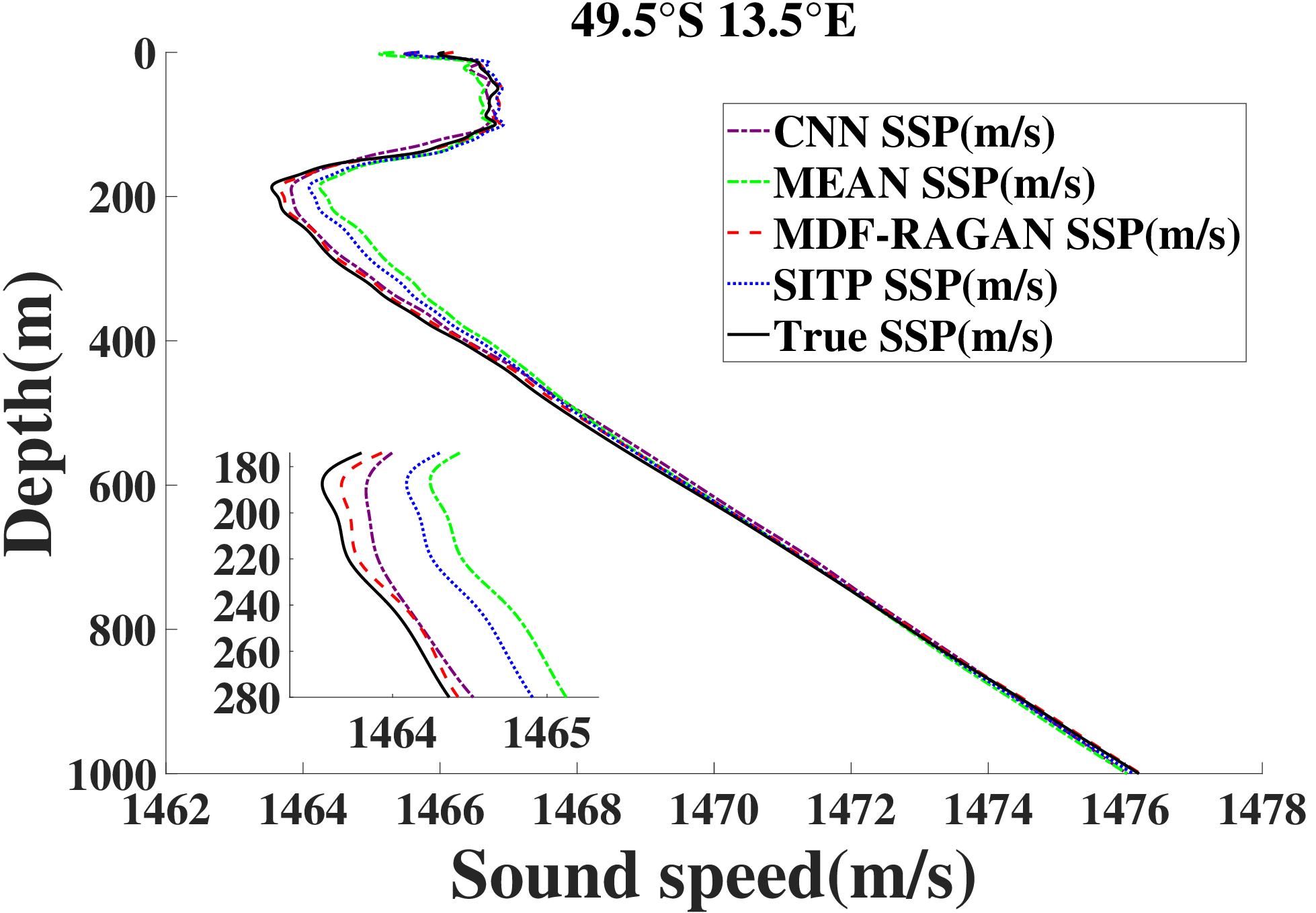}
		\caption{($49.5^{\circ}$S, $13.5^{\circ}$E), 1000m}
		\label{fig:ssp-comp-g}
	\end{subfigure}
	\begin{subfigure}[b]{0.32\textwidth}
		\includegraphics[width=\textwidth]{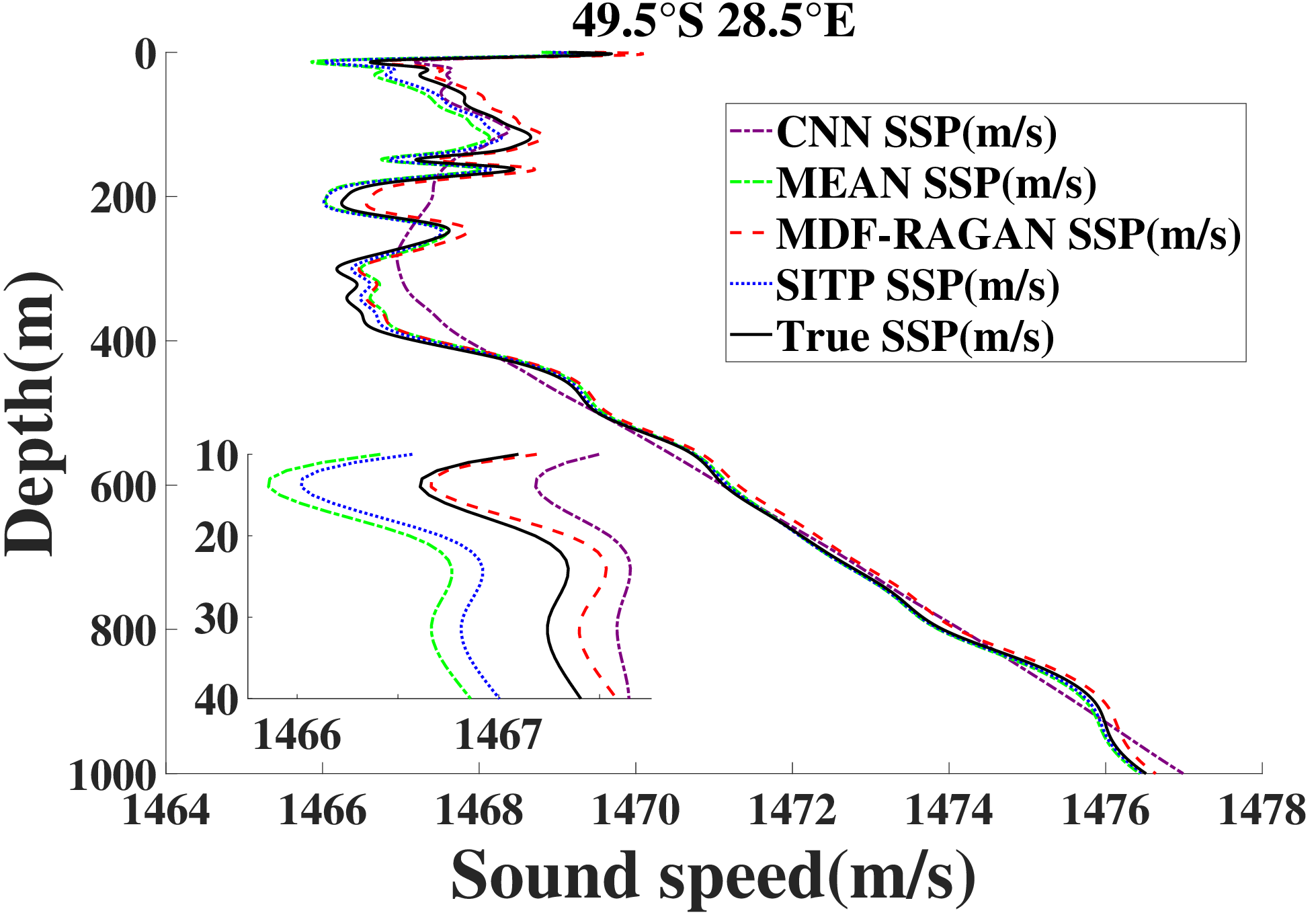}
		\caption{($49.5^{\circ}$S, $28.5^{\circ}$E), 1000m}
		\label{fig:ssp-comp-h}
	\end{subfigure}
	\begin{subfigure}[b]{0.32\textwidth}
		\includegraphics[width=\textwidth]{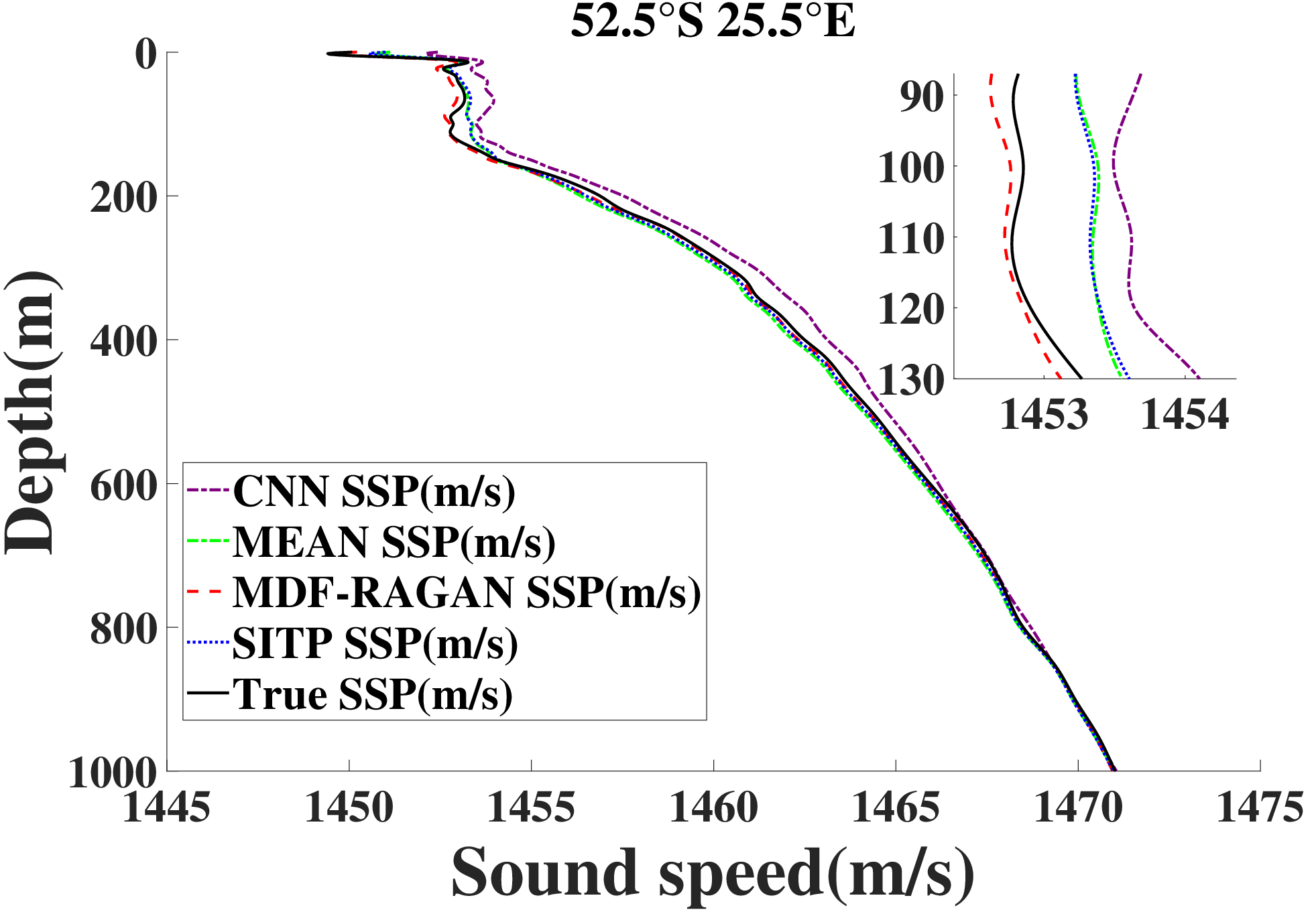}
		\caption{($52.5^{\circ}$S, $25.5^{\circ}$E), 1000m}
		\label{fig:ssp-comp-i}
	\end{subfigure}
	
	\begin{subfigure}[b]{0.32\textwidth}
		\includegraphics[width=\textwidth]{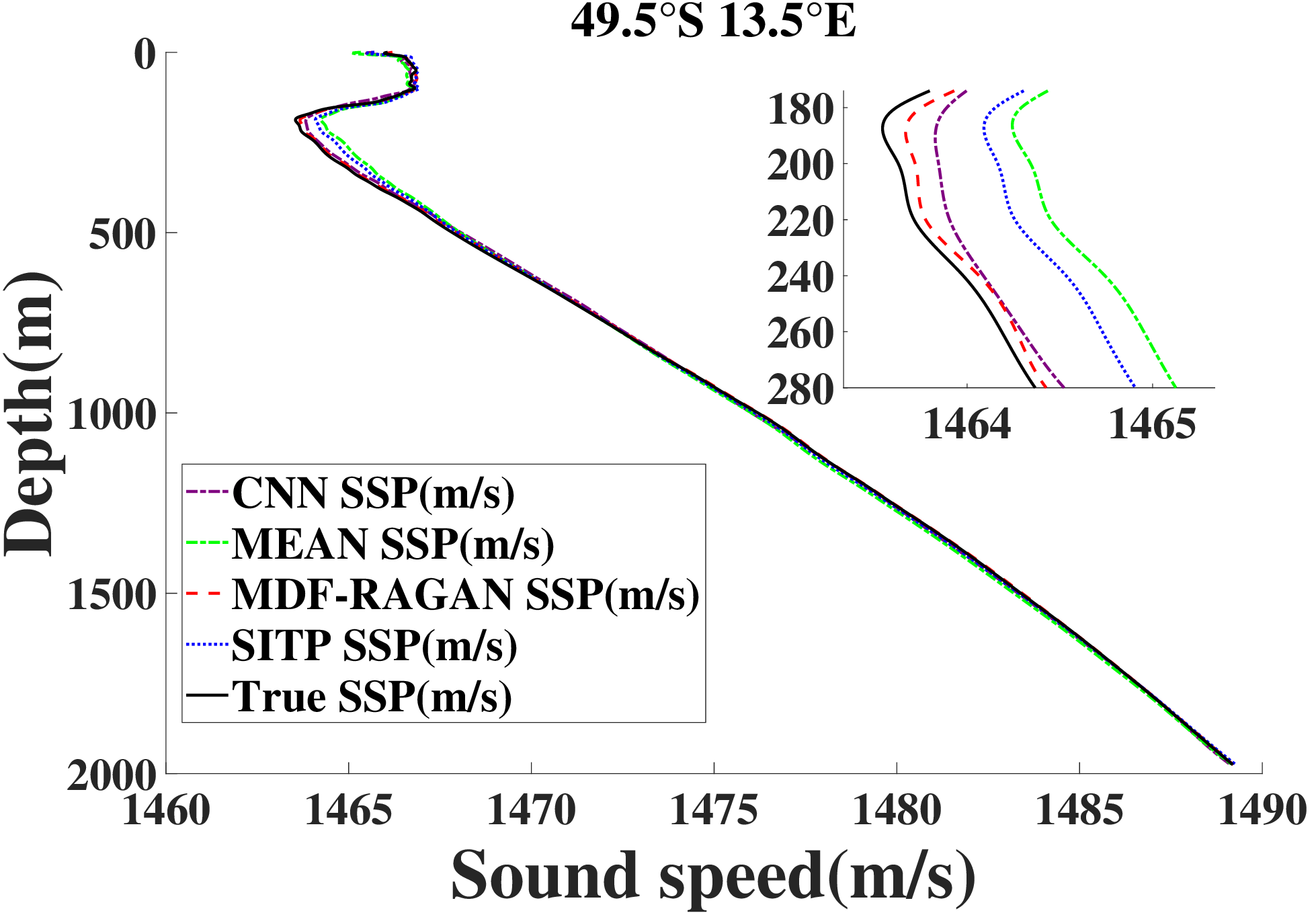}
		\caption{($49.5^{\circ}$S, $13.5^{\circ}$E), 1975m}
		\label{fig:ssp-comp-j}
	\end{subfigure}
	\begin{subfigure}[b]{0.32\textwidth}
		\includegraphics[width=\textwidth]{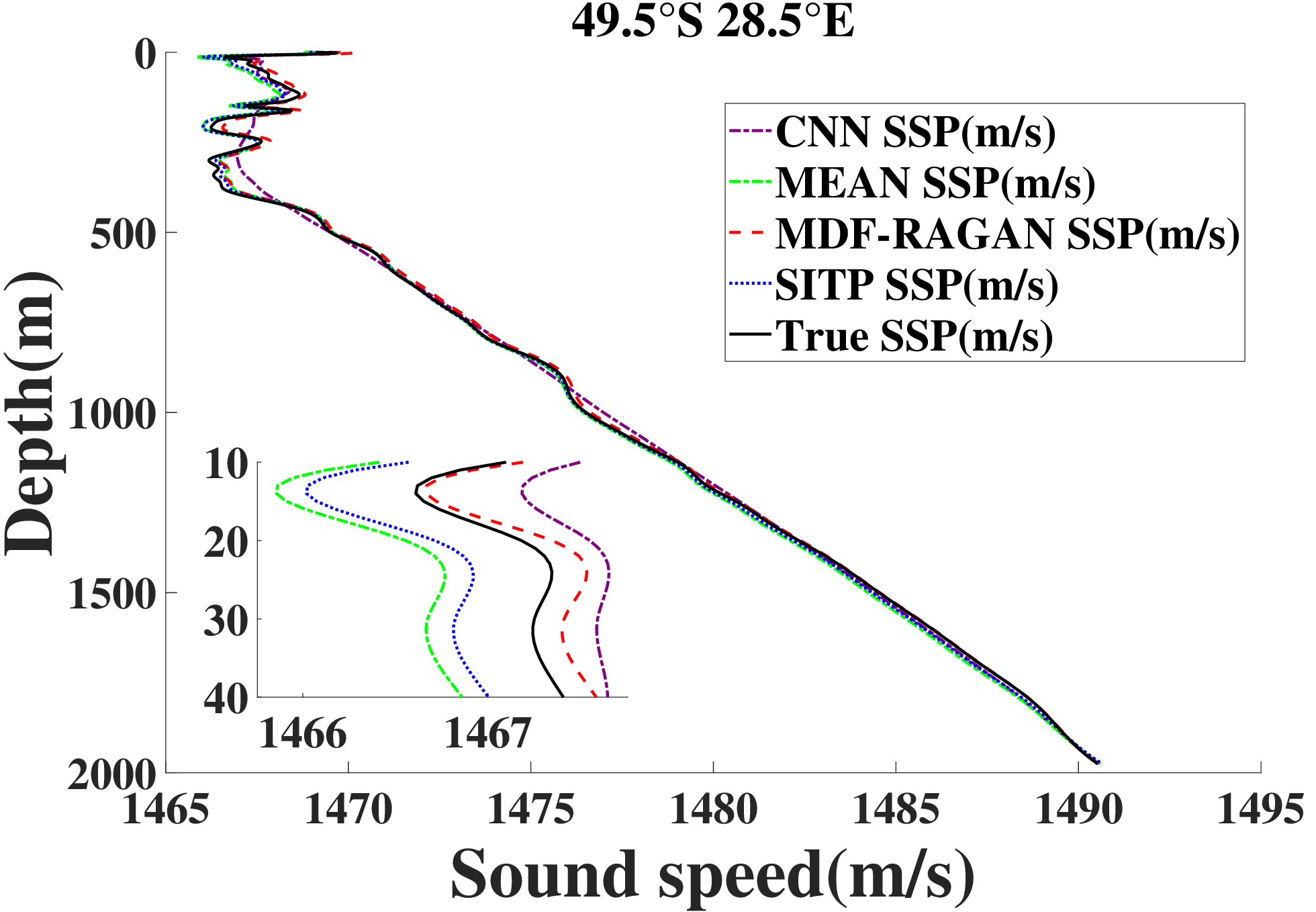}
		\caption{($49.5^{\circ}$S, $28.5^{\circ}$E), 1975m}
		\label{fig:ssp-comp-k}
	\end{subfigure}
	\begin{subfigure}[b]{0.32\textwidth}
		\includegraphics[width=\textwidth]{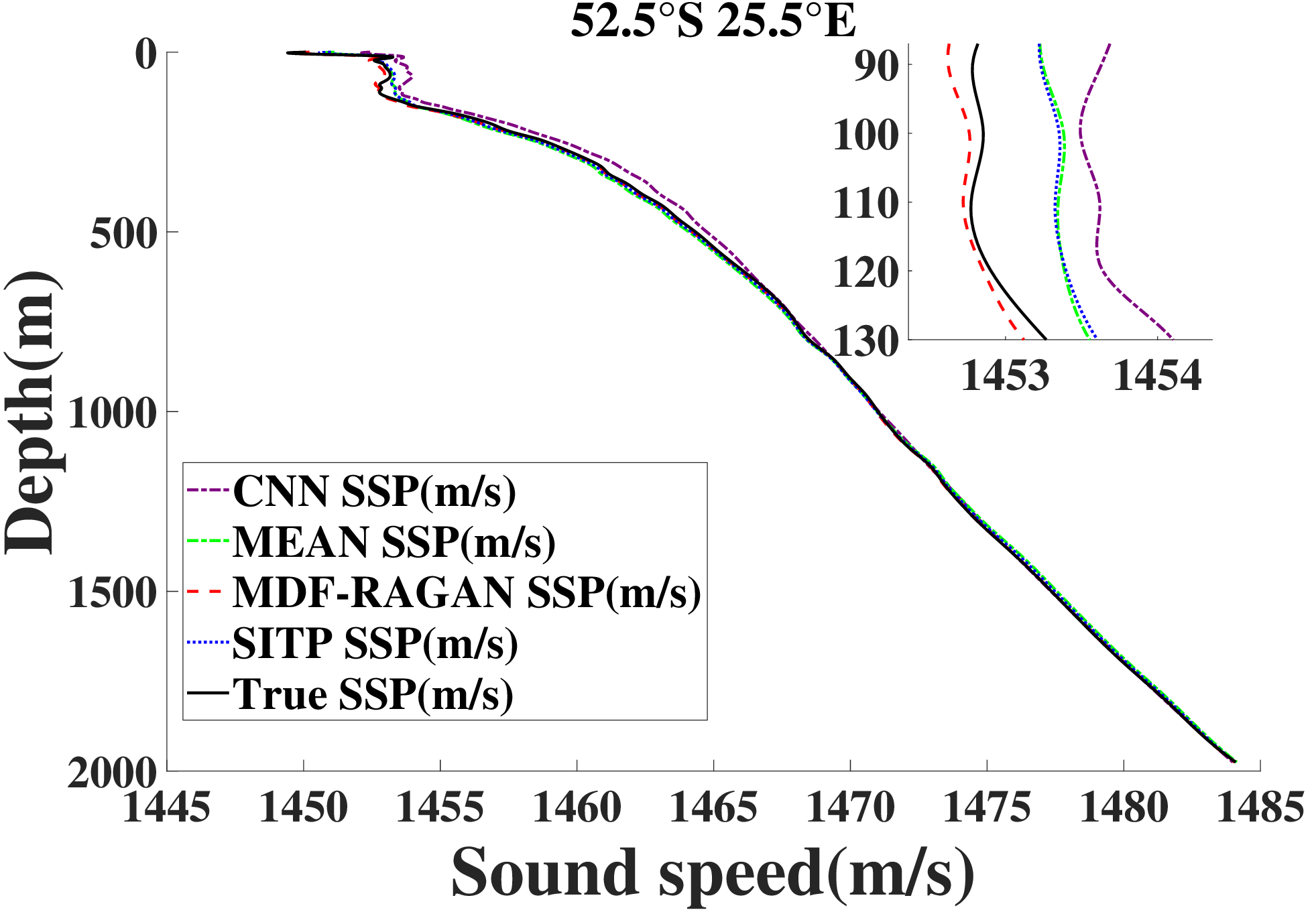}
		\caption{($52.5^{\circ}$S, $25.5^{\circ}$E), 1975m}
		\label{fig:ssp-comp-l}
	\end{subfigure}
	
	\caption{Comparison of sound speed profiles at different locations and depths. Each row represents a specific depth (200 m, 400 m, 1000 m, and 1975 m), while each column represents a specific location (Location 1: ($49.5^{\circ}$S, $13.5^{\circ}$E); Location 2: ($49.5^{\circ}$S, $28.5^{\circ}$E); Location 3: ($52.5^{\circ}$S, $25.5^{\circ}$E)). The figures illustrate the comparison of sound speed profile predictions using four different methods: MDF-RAGAN, CNN, SITP, and MEAN at each location and depth.}
	\label{fig04}
\end{figure*}

In contrast to the smooth sound velocity variations in the deep-sea region, the SSPs of shallow water bodies tend to show complex nonlinear fluctuations at the thermocline and scattering layer due to strong tidal, wind and internal wave action. Fig.~\ref{fig04} focuses on the estimation results of different methods at 200 m, 400 m, 1000 m, and 1975 m depths at selected shallow sea coordinates. At the 200 m depth section, the prediction curve of the historical averaging method is too flat to reflect the sharp rise of the thermocline due to the complete loss of spatial and temporal information; the inverse distance weighting method introduces the physical neighborhood of the peripheral points, but its fixed power weight limits the flexible response to the change of the thermocline amplitude, and thus the predicted peak is often overestimated or underestimated. The CNN is able to capture some of the mutations at 200 m, but it struggles to predict the exact peak values. At 400 m, the CNN can also capture some mutations, though it jitters before and after local extreme points.

The shallow sea profile enters the scattering layer at the stage of 200-1000 m. The sound velocity changes in this region are smoother, but there are still small oscillations. Although the errors of MEAN and SITP are slightly reduced, MDF-RAGAN is still much better than the other methods, and the prediction curves of CNN are alleviated at this layer, but it is still difficult to maintain the smoothness at the full depth. In contrast, MDF-RAGAN successfully smoothes out the high-frequency noise and closely follows the measured curves by virtue of the cross-modal attention that dynamically adjusts the fusion weights of the neighborhood and labeled features at each layer of depth. More detailed results at 4 typical locations are given in Table~\ref{tab05}.

\begin{figure}[!htbp]
	\centering
	\begin{subfigure}[b]{0.8\linewidth}
		\centering
		\includegraphics[width=\linewidth]{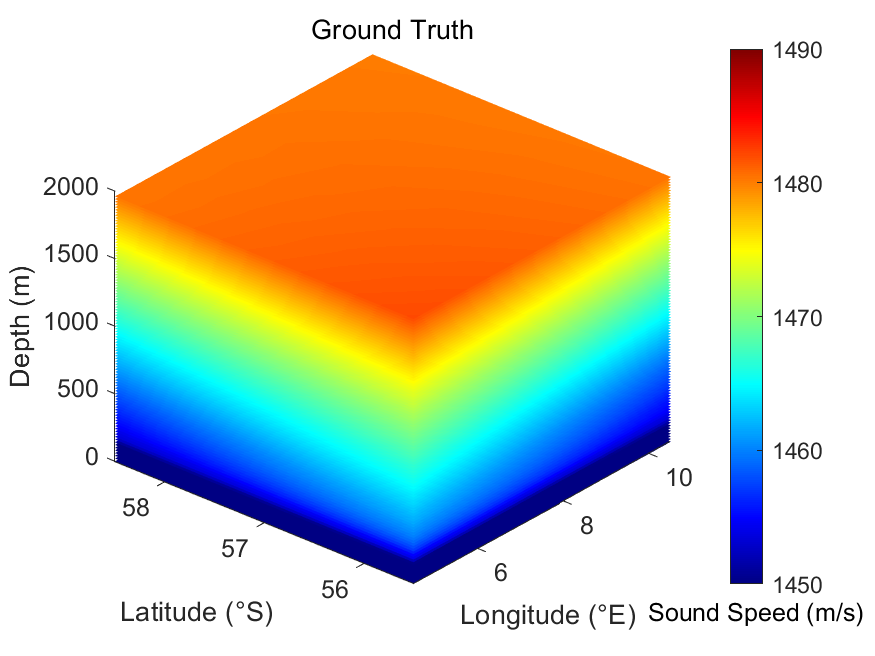}
		\caption{Ground truth}
	\end{subfigure}
	\\
	\begin{subfigure}[b]{0.8\linewidth}
		\centering
		\includegraphics[width=\linewidth]{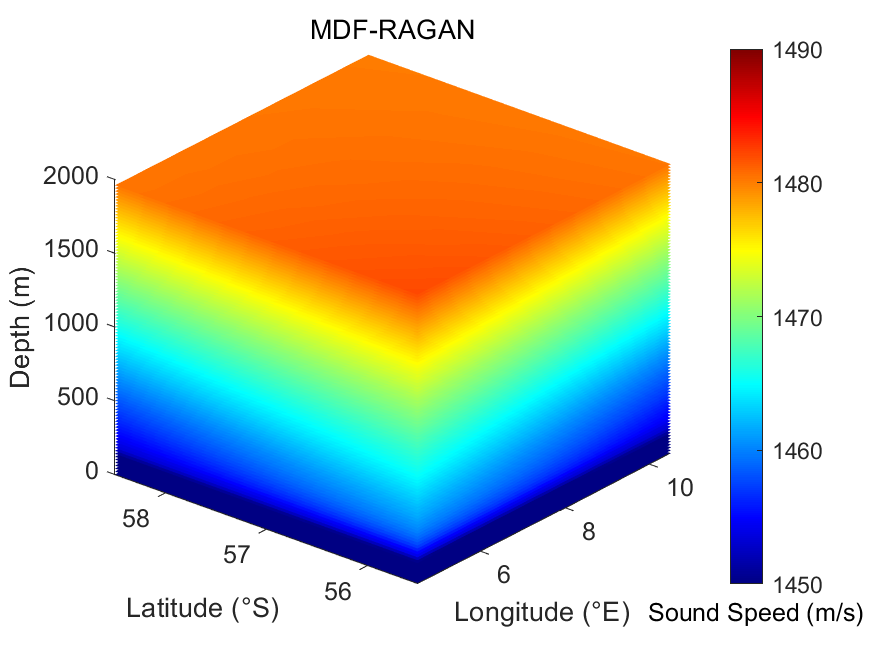}
		\caption{MDF-RAGAN}
	\end{subfigure}
	\caption{Predicted real-time 3D sound speed distribution. (a) Ground truth, (b) MDF-RAGAN.}
	\label{fig05}
\end{figure}

\begin{table}[htbp]
	\centering
	\fontsize{8pt}{10pt}\selectfont
	\caption{Comparison of RMSE in shallow sea areas at different depths}
	\label{tab05}
	\begin{tabular}{ccccc}
		\toprule
		\multirow{2}{*}{Method} & \multirow{2}{*}{Location} & \multicolumn{3}{c}{Depth (m)} \\
		\cmidrule(lr){3-5}
		& & 200 & 400 & 1000 \\
		\midrule
		\multirow{4}{*}{MDF-RAGAN} & ($49.5^{\circ}$S, $16.5^{\circ}$E) & 0.248 & 0.219 & 0.181 \\
		& ($49.5^{\circ}$S, $34.5^{\circ}$E) & 0.256 & 0.237 & 0.128 \\
		& ($49.5^{\circ}$S, $28.5^{\circ}$E) & 0.264 & 0.234 & 0.200 \\
		\midrule
		\multirow{4}{*}{Att-CNN} & ($49.5^{\circ}$S, $16.5^{\circ}$E) & 0.484 & 0.462 & 0.397 \\
		& ($49.5^{\circ}$S, $34.5^{\circ}$E) & 0.435 & 0.483 & 0.294 \\
		& ($49.5^{\circ}$S, $28.5^{\circ}$E) & 0.499 & 0.475 & 0.372 \\
		\midrule
		\multirow{4}{*}{MEAN} & ($49.5^{\circ}$S, $16.5^{\circ}$E) & 0.300 & 0.324 & 0.266 \\
		& ($49.5^{\circ}$S, $34.5^{\circ}$E) & 1.443 & 1.501 & 0.511 \\
		& ($49.5^{\circ}$S, $28.5^{\circ}$E) & 0.733 & 0.578 & 0.413 \\
		\midrule
		\multirow{4}{*}{SITP} & ($49.5^{\circ}$S, $16.5^{\circ}$E) & 0.255 & 0.236 & 0.185 \\
		& ($49.5^{\circ}$S, $34.5^{\circ}$E) & 1.082 & 1.122 & 0.460 \\
		& ($49.5^{\circ}$S, $28.5^{\circ}$E) & 0.520 & 0.419 & 0.292 \\
		\bottomrule
	\end{tabular}
\end{table}

Overall, MDF-RAGAN's unified modeling of abrupt and gradual sound speed changes in shallow sea environments is attributed to its multi-source fusion architecture: the LFB provides the coordinates and SST a prior that can constrain the anomaly weights, while the FMB stabilizes the gradients in the depth dimension, allowing the model to adaptively capture perturbed signals in both the thermocline and scattering layers, thus obtaining excellent profile prediction performance under the complex hydrodynamic conditions of shallow seas. The real-time 3D sound speed construction result is given in Fig.~\ref{fig05}, which was obtained by interpolating from the estimated results of 6 coordinates.

\subsubsection{Ablation Study}
\label{sec3-3-ablation}

To further quantify the contribution of each core component in MDF-RAGAN, we conducted an ablation study by progressively disabling different modules while keeping the remaining settings unchanged. Specifically, we considered four degraded variants: (i) \emph{w/o residual}, which removed the residual perturbation branch and directly regresses the full SSP instead of the perturbation around the neighbourhood mean profile; (ii) \emph{w/o FMB}, which discards the Feature Mapping Block and thus removes the depth-wise enhancement of SSP features; (iii) \emph{w/o SST}, which feeds only historical SSPs and geographic coordinates into the network while entirely removing the SST modality, thereby breaking the explicit coupling between surface temperature and subsurface sound speed; and (iv) \emph{w/o CMPAB}, which eliminates the Cross-Modal Perturbation Attention Block so that the model can no longer adaptively fuse labels and reference SSPs across large-scale spatial regions.

\begin{table}[!htbp]
	\centering
	\fontsize{8pt}{10pt}\selectfont
	\caption{Ablation study on the main components of MDF-RAGAN.}
	\label{tab:ablation}
	\begin{tabularx}{\linewidth}{cccccc}
		\toprule
		\textbf{Variant} & MDF-RAGAN & \makecell{w/o \\residual} & \makecell{w/o \\ FMB} & \makecell{w/o \\ SST} & \makecell{w/o \\ CMPAB} \\
		\midrule
		\textbf{RMSE (m/s)} &  0.148 & 1.006 & 0.256 & 0.641 & 0.675 \\
		\bottomrule
	\end{tabularx}
\end{table}

As reported in Table~\ref{tab:ablation}, removing the residual perturbation formulation (\emph{w/o residual}) severely degrades the reconstruction accuracy, with RMSE increasing from 0.148~m/s to 1.006~m/s, which confirms that learning fine-grained perturbations around the neighborhood mean profile is much easier than regressing the entire SSP from scratch. The performance drops observed in the \emph{w/o FMB} and \emph{w/o CMPAB} variants further demonstrate that both depth-wise feature enhancement and cross-modal perturbation attention are indispensable for capturing thermocline structures and long-range spatial dependencies. In particular, the \emph{w/o SST} variant yields a RMSE of 0.641~m/s, highlighting that explicitly incorporating SST as an additional modality provides crucial physical information on upper-ocean thermal gradients, and plays a key role in improving the fidelity of SSP prediction. From a physical perspective, SST controls the upper-ocean thermohaline gradient and thus provides a strong prior for the vertical structure of the SSP, especially around the thermocline. By explicitly incorporating SST as a remote-sensing modality, MDF-RAGAN can exploit globally available satellite products to improve SSP estimation in data-sparse regions where in-situ profiles are scarce.

\subsection{Discussions}

\begin{figure*}[!htbp]
	\centering
	\begin{subfigure}[b]{0.3\textwidth}
		\centering
		\includegraphics[width=\textwidth]{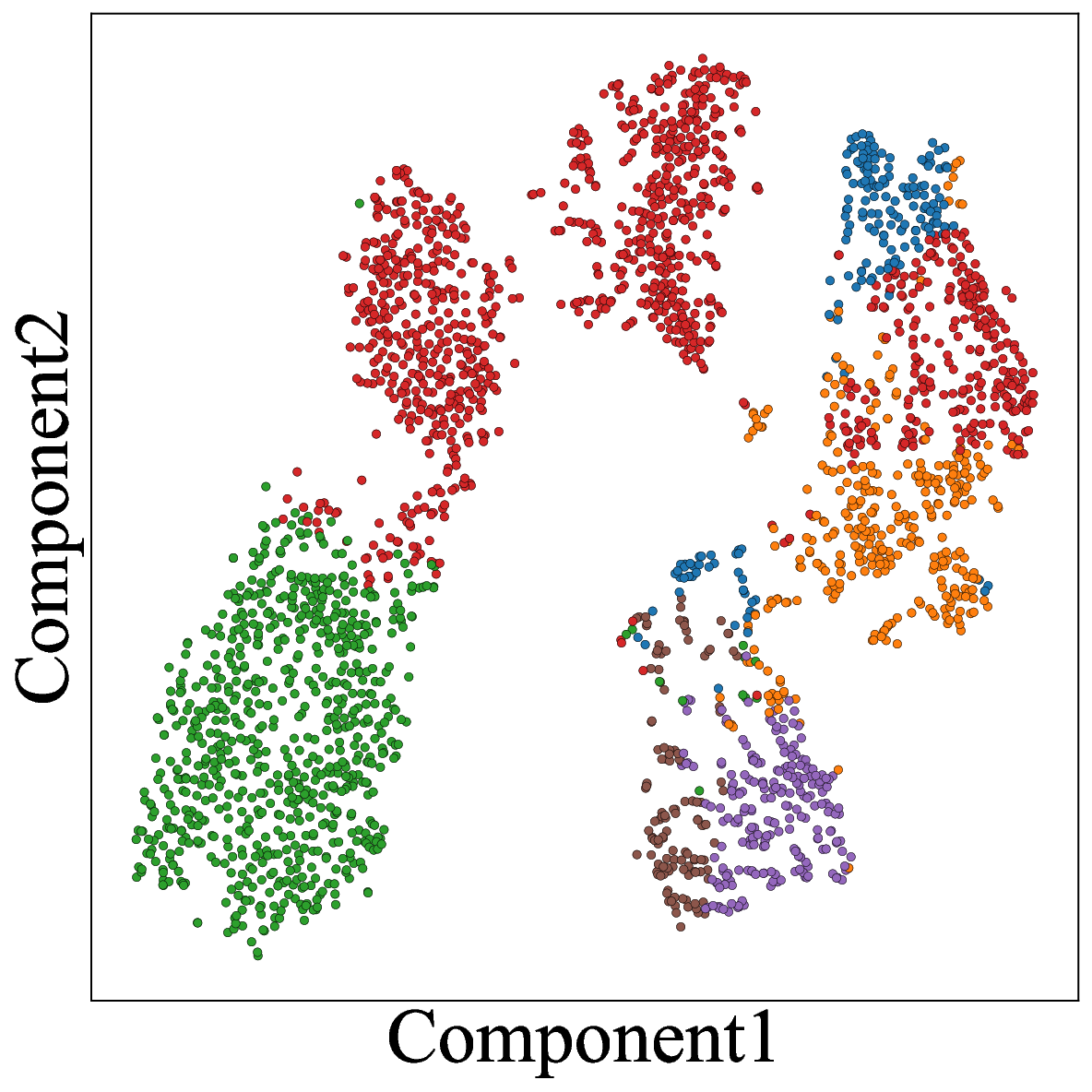}
		\caption{}
		\label{fig:tsne-a}
	\end{subfigure}
	\hfill
	\begin{subfigure}[b]{0.3\textwidth}
		\centering
		\includegraphics[width=\textwidth]{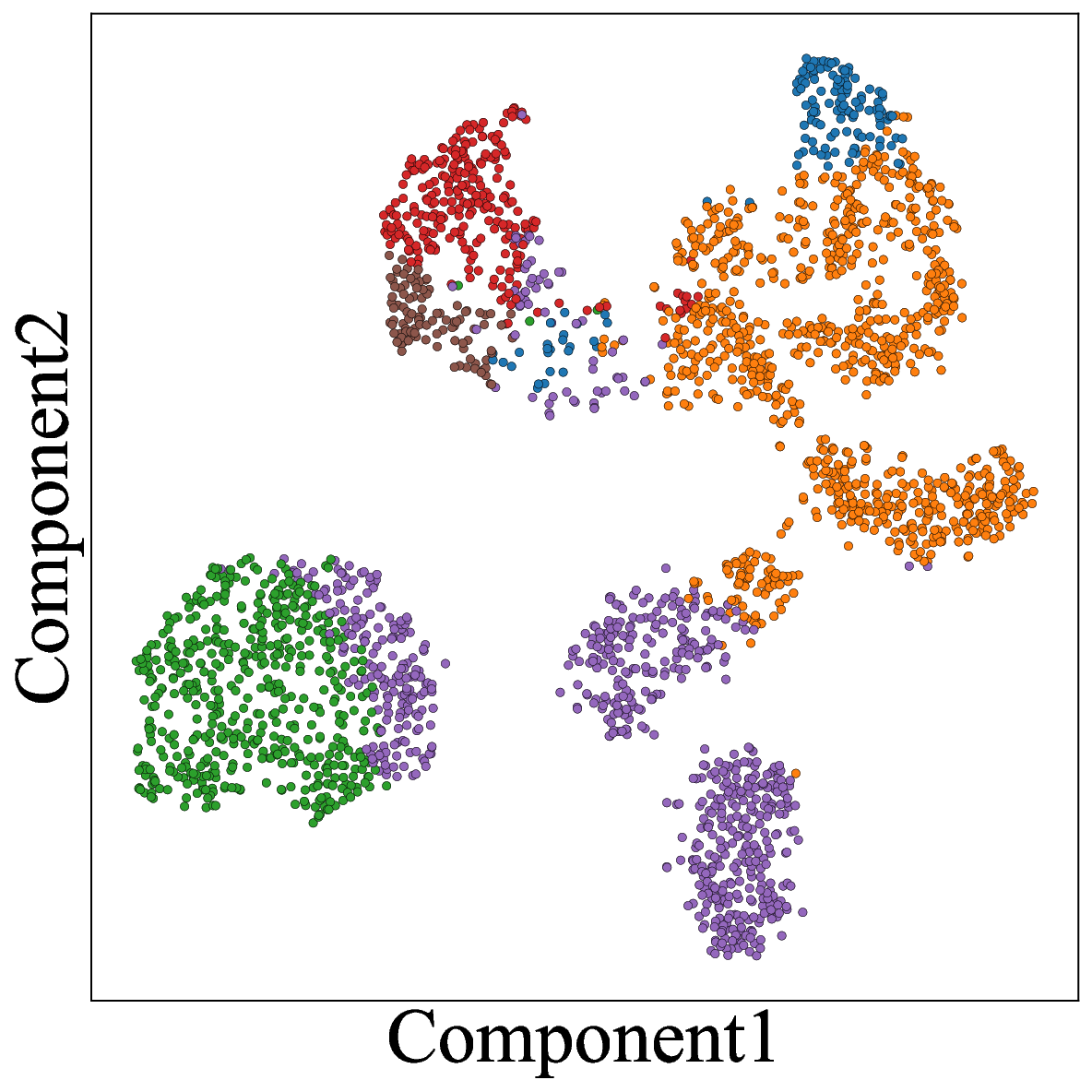}
		\caption{}
		\label{fig:tsne-b}
	\end{subfigure}
	\hfill
	\begin{subfigure}[b]{0.3\textwidth}
		\centering
		\includegraphics[width=\textwidth]{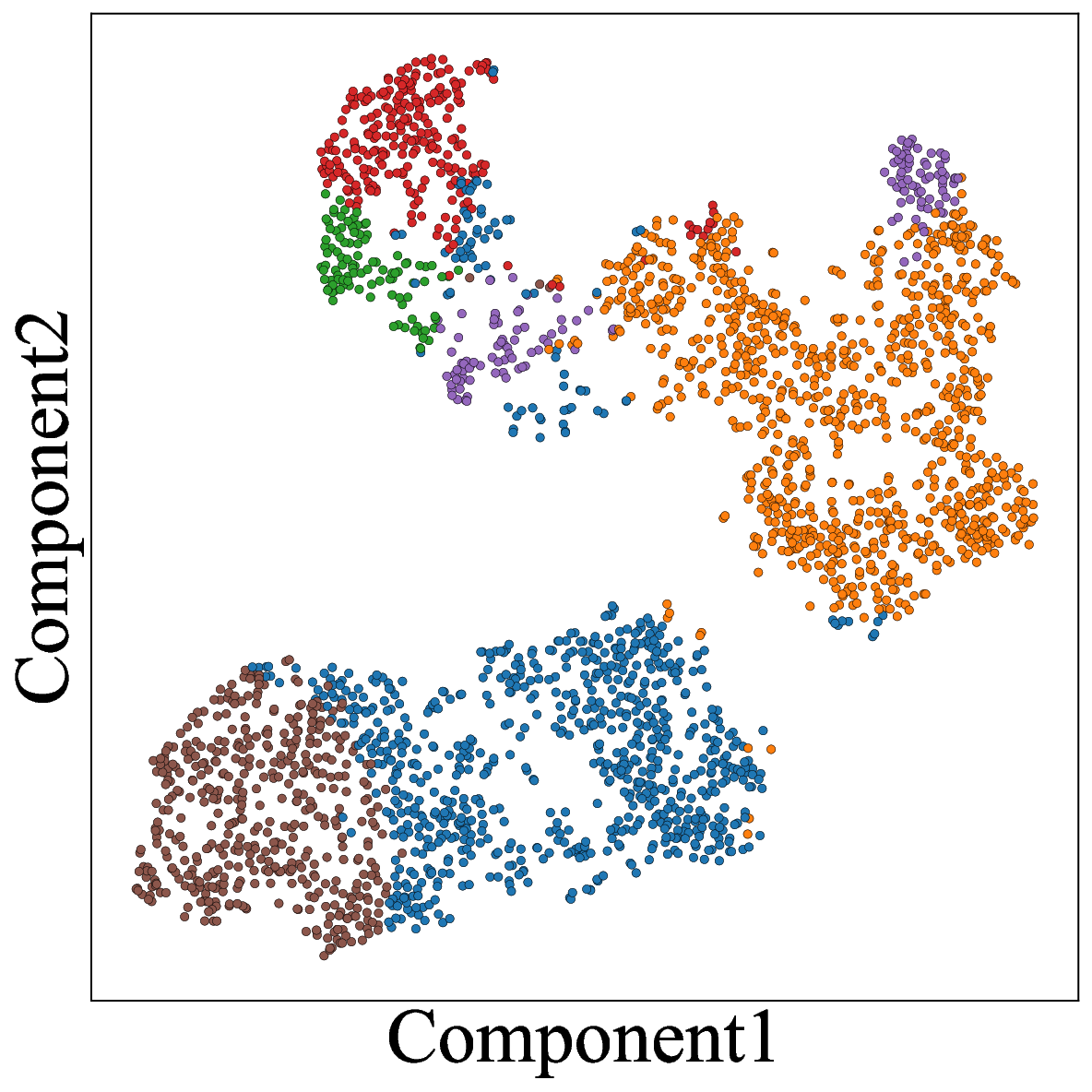}
		\caption{}
		\label{fig:tsne-c}
	\end{subfigure}
	\caption{t-SNE visualization of intermediate features in MDF-RAGAN. (a)-(c) show the intermediate features from the discriminator: (a) neighbor label features, (b) target SST-label features, and (c) target LOC-label features.}
	\label{fig06}
\end{figure*}

To evaluate the reconstruction capability of MDF-RAGAN in the feature space, we extracted three types of intermediate embedding features from the discriminator—neighbourhood labels (N-Label), SST labels (T-Label), and coordinate labels (LOC-Label)—and visualised them using t-SNE (Fig.\ref{fig05}). Fig.\ref{fig:tsne-a} shows that the true and generated N-Label samples form clearly separated clusters in the projection space, indicating that multi-source fusion significantly enhances the discriminator's ability to identify neighbourhood perturbations. The distribution of T-Label features in Fig.\ref{fig:tsne-b} also exhibits good discriminability, further validating the constraining role of physical priors in the SST channel during adversarial learning. In contrast, the generated distribution of LOC-Label features shown in Fig.\ref{fig:tsne-c} is almost completely nested within the real sample cluster, indicating that the model has high accuracy and consistency in reconstructing the label space distribution of target points. The above results collectively demonstrate that the cross-modal deep fusion mechanism implemented by the CMPAB not only enhances the discriminative power of feature representation but also ensures the interpretability and credibility of generated samples in terms of physical semantics. Overall, the distribution structures of the three types of intermediate features systematically validate the effectiveness of MDF-RAGAN in multi-source alignment and local perturbation modelling, providing robust theoretical support and empirical evidence for its application in predicting sound speed profiles in complex marine environments.

Although t-SNE visualisation has validated the physical consistency and semantic interpretability of intermediate features in the representation space, further investigation into the contribution of these learned representations to actual prediction performance remains critical. To this end, we employ empirical cumulative distribution function (ECDF) analysis to assess the absolute prediction error of different methods at various ocean depths. Specifically, we introduce the empirical cumulative distribution function (ECDF) as an evaluation metric to characterise the absolute error distribution of each method at different depths. For each depth layer, we calculate the absolute error between the predicted values and the true values of all samples at that depth point, and plot the cumulative distribution curve of the error magnitude based on this. The horizontal axis of the ECDF curve represents the magnitude of the error (in millimetres), while the vertical axis represents the proportion of samples with errors not exceeding that value.

\begin{figure*}[htbp]
	\centering
	\begin{subfigure}[b]{0.4\textwidth}
		\includegraphics[width=\textwidth]{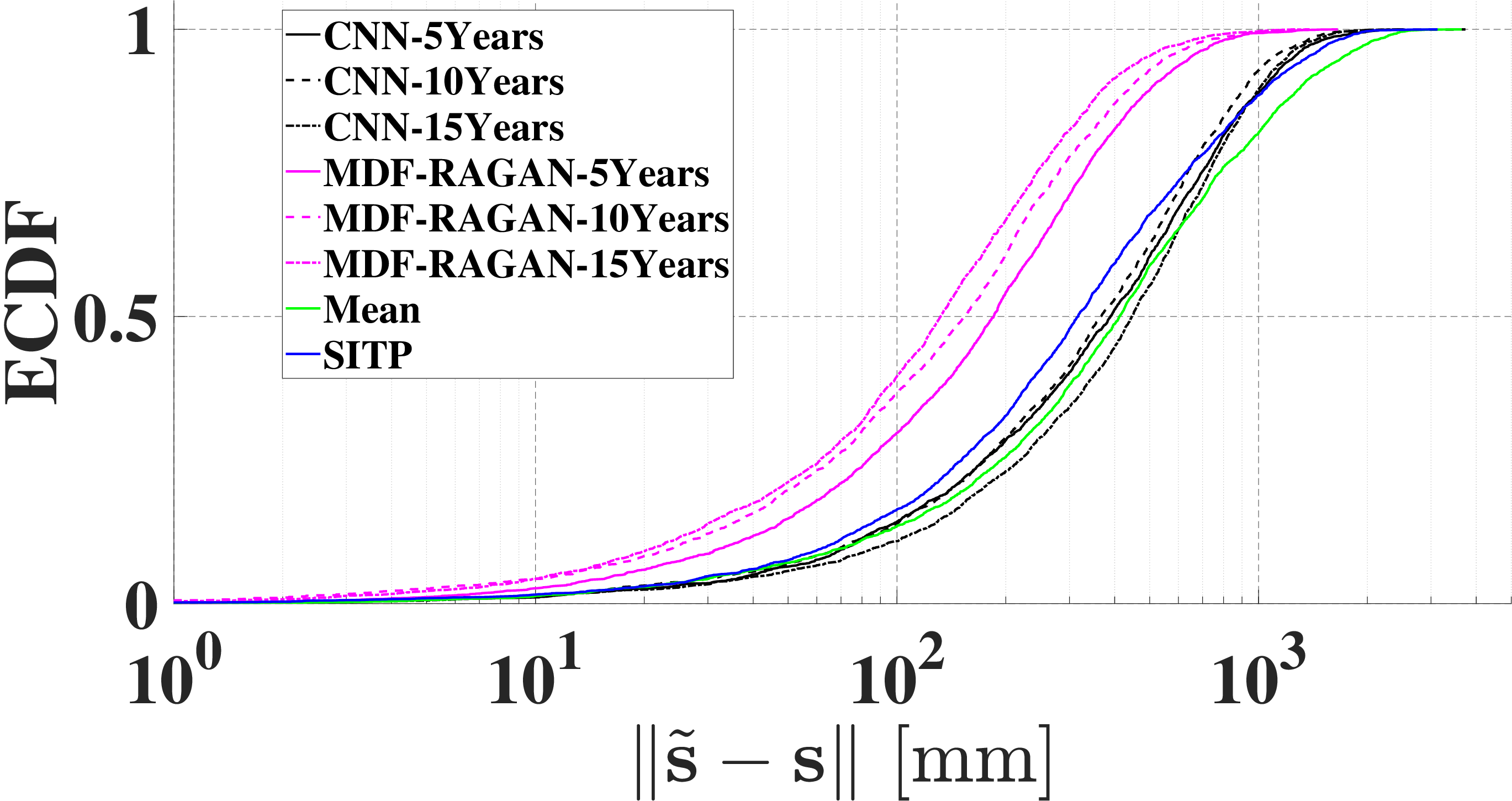}
		\caption{200m}
	\end{subfigure}
	\begin{subfigure}[b]{0.4\textwidth}
		\includegraphics[width=\textwidth]{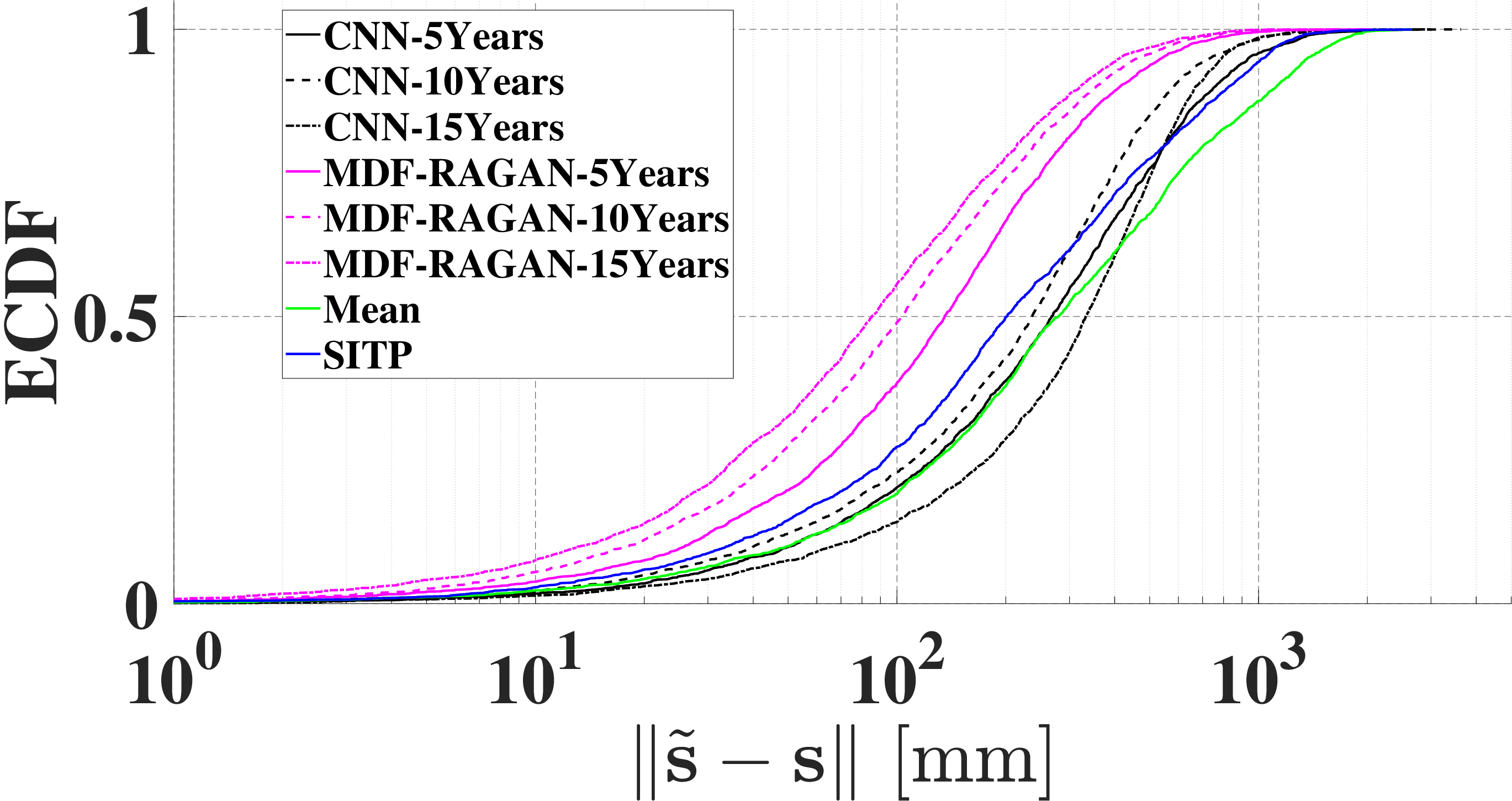}
		\caption{400m}
	\end{subfigure}
	
	\begin{subfigure}[b]{0.4\textwidth}
		\includegraphics[width=\textwidth]{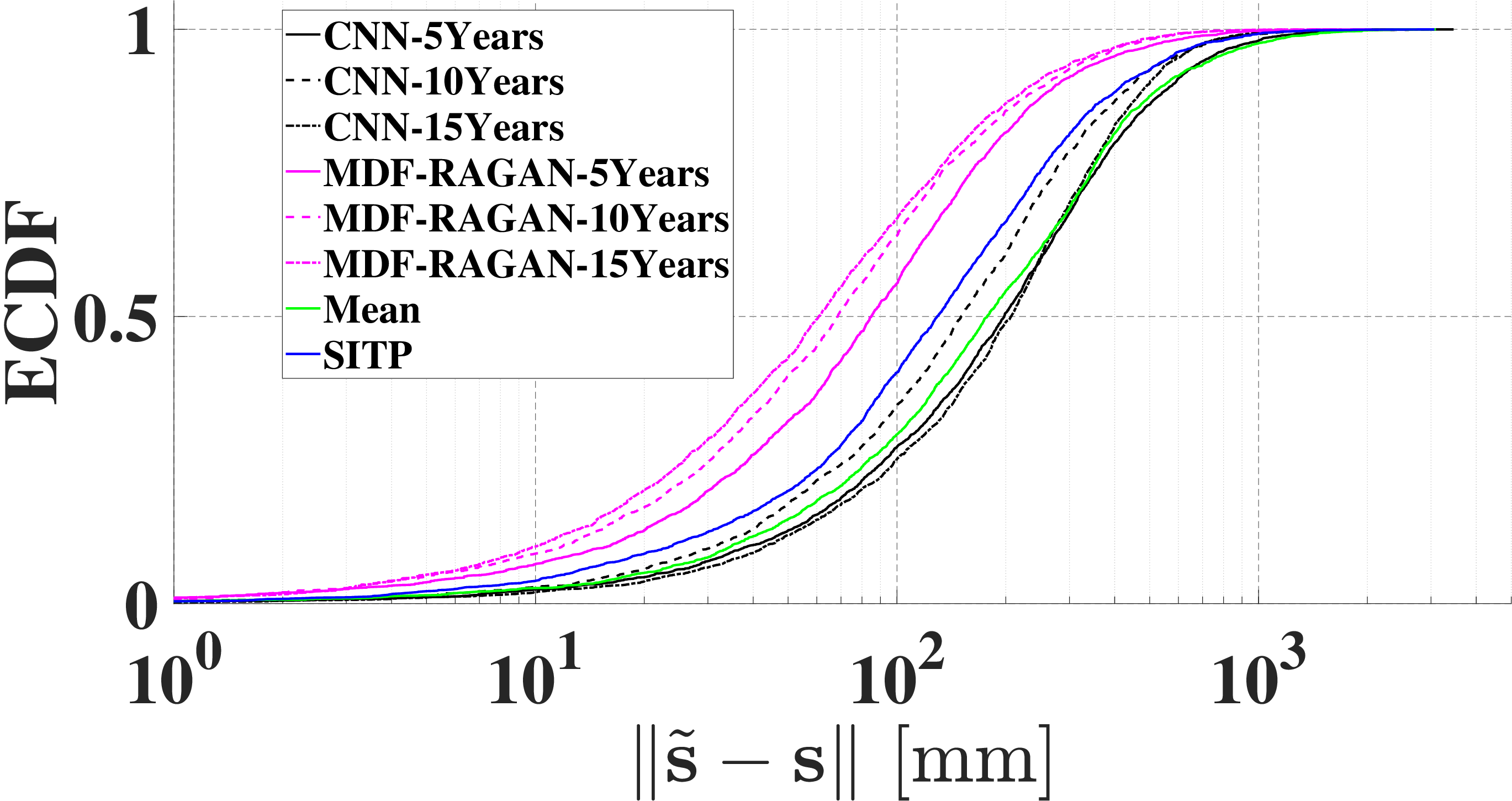}
		\caption{800m}
	\end{subfigure}
	\begin{subfigure}[b]{0.4\textwidth}
		\includegraphics[width=\textwidth]{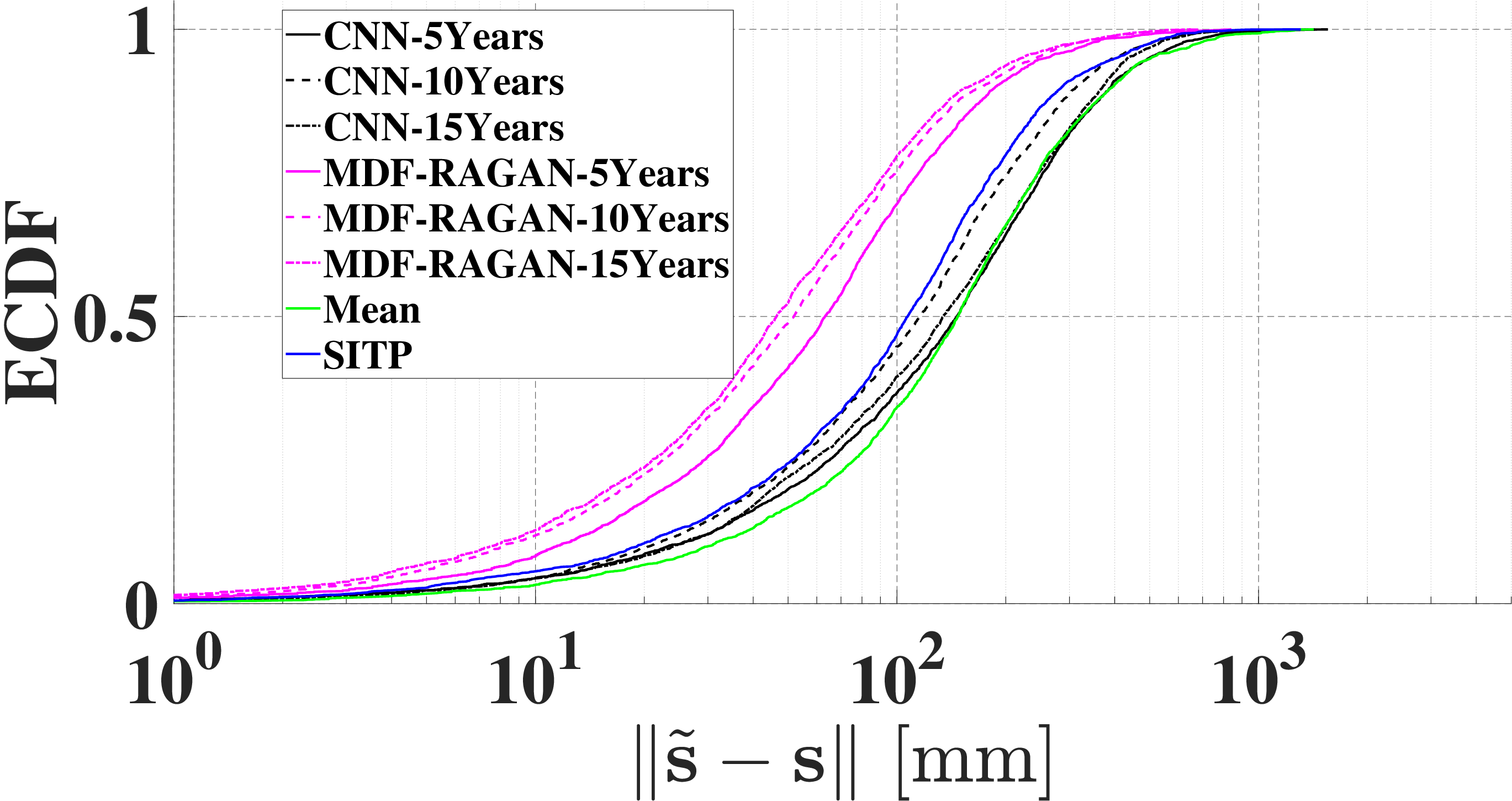}
		\caption{1000m}
	\end{subfigure}
	
	\begin{subfigure}[b]{0.4\textwidth}
		\includegraphics[width=\textwidth]{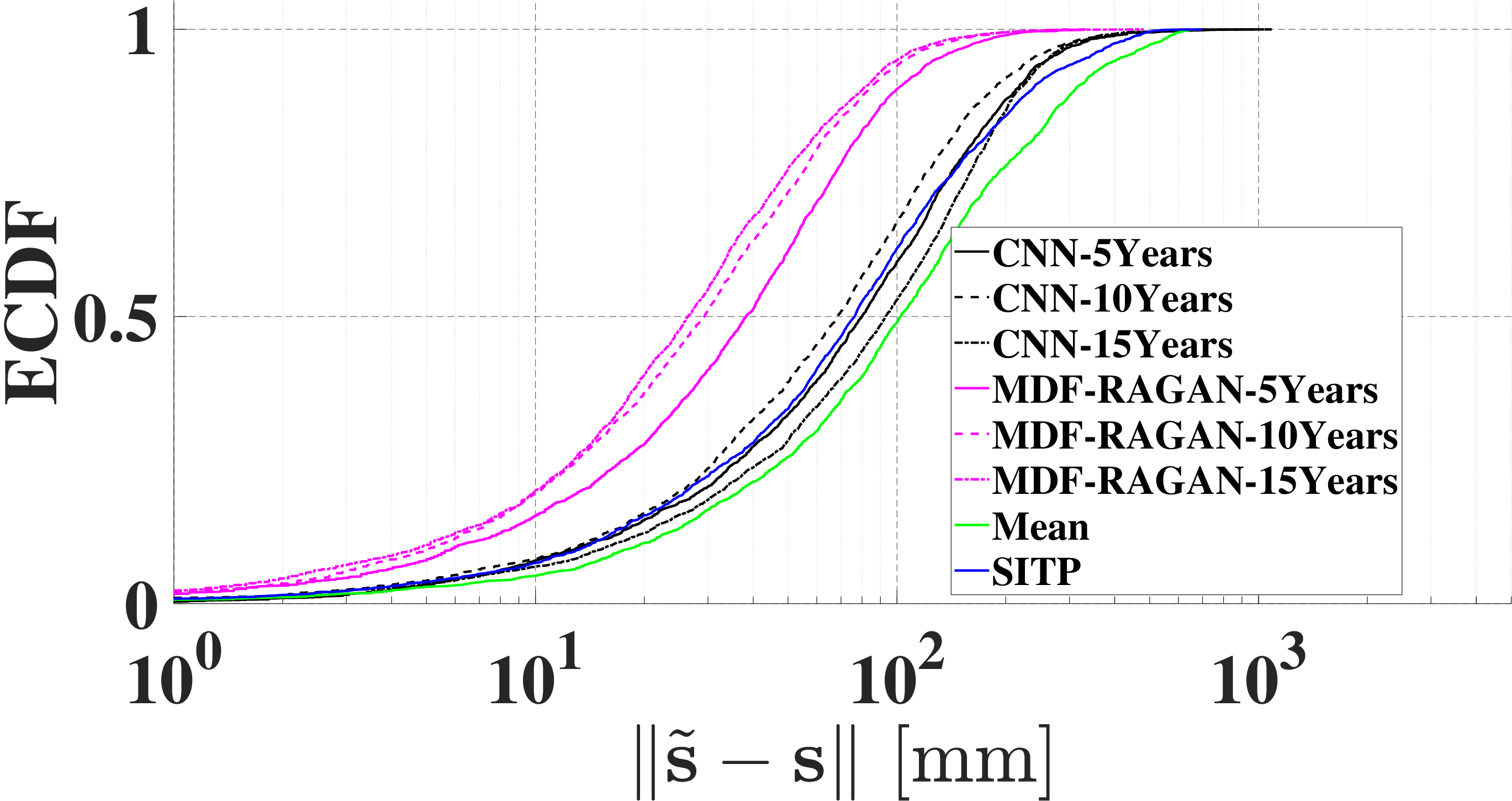}
		\caption{1400m}
	\end{subfigure}
	\begin{subfigure}[b]{0.4\textwidth}
		\includegraphics[width=\textwidth]{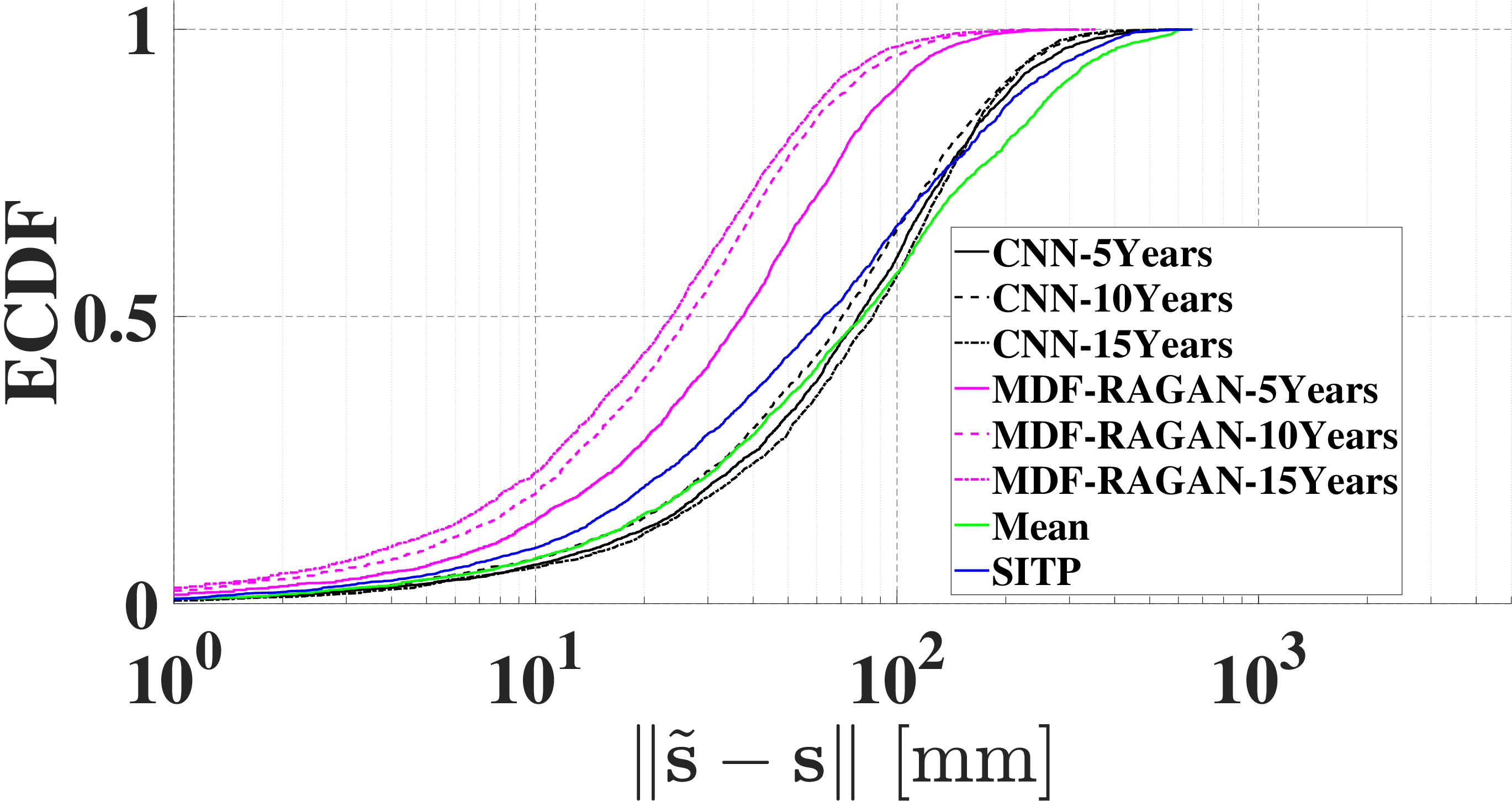}
		\caption{1975m}
	\end{subfigure}
	
	\caption{
		ECDF comparison of absolute prediction errors for different methods at various depths. (a) 200m, (b) 400m, (c) 800m, (d) 1000m, (e) 1400m, (f) 1975m. The curves show that MDF-RAGAN consistently achieves smaller errors and outperforms other baselines across all tested depths.
	}
	\label{fig07}
\end{figure*}

\begin{table*}[htbp]
	\centering
	\fontsize{8pt}{10pt}\selectfont
	\caption{Comparison of overall RMSE (m/s), model size, inference time and training time across different methods}
	\label{tab07}
	{
		\setlength{\tabcolsep}{2pt}
		\begin{tabular}{lcccc}
			\toprule
			\textbf{Methods} & \textbf{RMSE (m/s)} & \textbf{Parameters} & \textbf{Inference Time (ms)} & \textbf{Training Time} \\
			\midrule
			MDF-RAGAN & 0.148 & 7.7M & 4.48 & 1 h 18 min 2 s \\
			Att-CNN & 0.311 & 4.69M & 0.87 & 38 min 49 s \\
			SITP & 0.338 & - & - & - \\
			MEAN & 0.434 & - & - & - \\
			\bottomrule
		\end{tabular}
	}
\end{table*}

Firstly, the ECDF curves show the differences in the cumulative distribution of errors between the methods at six typical depth layers (200, 400, 800, 1000, 1400, 1975 m). The ECDF curves for MDF-RAGAN are the most left-shifted and steepest rising in the shallow thermocline layer (200 m vs. 400 m, see Fig. \ref{fig06}a-b): at 200 m depth, about 80\% of the samples have an error of less than 50 mm, whereas CNN, SITP, and the mean baseline need to raise the threshold to 200-300 mm to achieve the same hit rate; at 400 m, MDF-RAGAN keeps 90\% of the quantile errors within 80 mm, far exceeding the 150-250 mm of the other methods.

Into the middle layer (800 m vs. 1000 m, Fig. \ref{fig06}c-d), the error spectra of all the methods are shifted to the left, but MDF-RAGAN remains ahead of CNN and SITP/MEAN at 80-100 mm with more than 50\% of sample errors below 30 mm. In the deeper layers (1400 m vs. 1975 m, Fig. \ref{fig06}e-f), the errors converge further, with MDF-RAGAN achieving a median error of about 20 mm at 1975 m, while CNN is still around 40 mm, and SITP/MEAN is as high as 80-100 mm. 

Comparison of the same-coloured solid, dashed and dotted lines in different historical windows (5, 10, 15 years) shows that MDF-RAGAN is the most sensitive to the expansion of the training set, with the ECDF curve shifted left by 10-20 mm. CNN shows only a small improvement, while SITP/MEAN shows few changes, highlighting the advantages of cross-modal attention in enhancing the learning efficiency and generalisation ability of scarce samples.

The comprehensive comparison results based on the whole-profile mean sound speed error (RMSE) are summarized in Table \ref{tab07}. It can be seen that MDF-RAGAN not only outperforms CNN and SITP by nearly a factor of two, but also achieves about 65.8\% RMSE reduction compared to MEAN, which fully reflects the enhancement of overall profile matching by multi-source fusion and cross-modal attention. By comparing the scale of model parameters and the execution time, our proposed model occupies more storage and requires more computation time. However, considering that training can be completed offline in advance and the execution time is still in milliseconds, it is acceptable in practical applications.

In the shallow sea error analysis, MDF-RAGAN shows lower and more stable RMSE at different depths and spatial locations, with the error fluctuation range of 0.095-0.372 m/s, while CNN and SITP show larger fluctuation in the shallow sea area. Even in the deep sea area where there is a slight convergence, their RMSE remains higher than 0.30 m/s, which is 110\% higher than MDF-RAGAN. SITP is limited by the fixed distance weighting and is not adaptive to the dynamic change of depth, with its RMSE being 128\% higher. MEAN completely ignores spatial and temporal information and has the largest deviation across the full-profile, with RMSE close to 0.44 m/s. In summary, MDF-RAGAN achieves a comprehensive lead in both point-to-point and distribution-level evaluation metrics, reflecting the comprehensive superiority of the proposed model in the sound speed profile prediction task.

\section{Conclusion}
\label{sec4}
To realize high-precision estimation of sound velocity distribution in a given sea area without on-site underwater data measurement, we propose the MDF-RAGAN model for SSP construction. The attention mechanism is embedded to improve the model's ability for capturing global spatial feature correlations, and the residual module is embedded for deeply capturing small disturbances in the deep ocean sound velocity distribution caused by changes of SST. Through experiments on real open dataset, it shows that the proposed model outperforms other state-of-the-art methods. Specifically, MDF-RAGAN not only outperforms CNN and SITP by nearly a factor of two, but also achieves about 65.8\% RMSE reduction compared to MEAN, which fully reflects the enhancement of overall profile matching by multi-source fusion and cross-modal attention.

\section*{CRediT authorship contribution statement}
\textbf{Wei Huang:} Conceptualization, Methodology, Writing - Original Draft, Writing - Review \& Editing,  Funding acquisition. \textbf{Yuqiang Huang:} Methodology, Software, Writing - Original Draft. \textbf{Jixuan Zhou:}  Writing - Review \& Editing, Project administration, Funding acquisition. \textbf{Hao Zhang:} Writing - Review \& Editing, Project administration, Funding acquisition.  \textbf{Tianhe Xu:} Writing - Review \& Editing, Project administration, Funding acquisition. \textbf{Qian Sun:} Writing - Review \& Editing, Project administration. \textbf{Fang Ji:} Writing - Review \& Editing, Project administration.

\section*{Data statement}
SSP data were taken from the GDCSM\_Argo gridded Argo dataset \citep{GDCSM}, and the SST data were obtained from the NOAA OISST monthly average product \citep{SST2021}. Data will be available on request.

\section*{Declaration of competing interest}
The authors declare that they have no known competing financial interests or personal relationships that could have appeared to influence the work reported in this paper.

\section*{Funding}
This work was supported in part by the National Key Research and Development Program of China under Grant 2024YFB3909701, in part by the National Natural Science Foundation of China under Grant 42404001 and 62271459, in part by the Frontier Exploration Project of Hanjiang National Laboratory under Grant TS2024026, in part by the Stable Supporting Fund of Acoustic Science and Technology Laboratory under Grant JCKYS2025SSJS008, and in part by the Fundamental Research Funds for the Central Universities, Ocean University of China under Grant 202562022.

\bibliographystyle{unsrt} 
\bibliography{draft.bib}



\end{document}